\documentclass[colorlinks=true,urlcolor=urlblue,linkcolor=black,citecolor=black,breaklinks=true,bookmarks=false,colorlinks=true, allcolors=blue]{optica-article}
\usepackage{hyperref}
\journal{opticajournal} 
\articletype{Research Article}
\usepackage{nicefrac}
\usepackage{amsmath}
\usepackage{graphicx}
\usepackage{array}
\usepackage{multirow}
\usepackage{makecell} 
\usepackage{cite}
\usepackage{subcaption}
\usepackage{siunitx}
\usepackage{tabularx}
\usepackage{comment}
\usepackage{color} 
\usepackage{textcomp} 
\usepackage{booktabs} 	
\usepackage{longtable}	
\usepackage{float}      
\usepackage[clean]{svg-extract}
\usepackage{lscape} 
\usepackage{mwe} 
\usepackage[percent]{overpic}
\usepackage{multirow}
\usepackage[normalem]{ulem}

\begin{document}
\title{Temperature-compensating package for OH line filters for astronomy: II. manufacture, assembly, and performance study}

\author{Xijie Luo,\authormark{1,2,*,\textdagger} Carlos E. Rodriguez Alvarez,\authormark{1,3,\textdagger} Aashia Rahman,\authormark{1} Azlizan A. Soemitro,\authormark{1,2} Hakan \"O{}nel,\authormark{1} Jens Paschke,\authormark{1} Svend-Marian Bauer,\authormark{1} Kalaga Madhav,\authormark{1,2} Wilbert Bittner,\authormark{1} and Martin M. Roth\authormark{1}}

\address{\authormark{\textdagger{} denotes equal contribution}\\
\authormark{1} Leibniz Institute for Astrophysics Potsdam (AIP), An der Sternwarte 16, 14482 Potsdam, Germany\\
\authormark{2} Institut für Physik und Astronomie, Universit\"a{}t Potsdam, Karl-Liebknecht-Str. 24/25, 14476 Potsdam, Germany\\
\authormark{3} Technische Universit\"a{}t Berlin, Fakult\"a{}t V- Institut für Mechanik, FG Strukturmechanik und
Strukturberechnung, Sekr. C 8-3, Geb. M, Straße des 17. Juni 135, 10623 Berlin, Germany}

\email{\authormark{*} xluo@aip.de} 

\begin{abstract*}
Multi-channel aperiodic fiber Bragg grating (FBG) based hydroxyl (OH) line filters have attracted significant interest in ground-based near-infrared (NIR) astronomical observations. In this paper, we present the performance of a new self-compensating enclosure for the filters, that can be used in non-temperature-controlled environments. 
Our prototype encloses a \SI{110}{mm} long single-mode photosensitive optical fiber with three  $\sim$\SI{10}{mm} filter gratings. A fourth grating was used as a reference outside the package to measure the uncompensated  wavelength shift. The prototype was tested over three thermal cycles, and showed a maximum wavelength deviation of \SI{12}{pm}, a wavelength drift of only \SI{0.37}{{pm}/{^{\circ}C}}, over the temperature range of  \SI{-17}{^{\circ}C} to \SI{15}{^{\circ}C}. The athermalization factor, i.e., the ratio of the maximum wavelength shift of the compensated grating to the uncompensated reference filter grating was \nicefrac{1}{22}. The results demonstrate the capability of the prototype for stabilizing multi-channel long-length FBGs or chirped FBGs, particularly for astronomical applications that require sub-picometer stability.

\end{abstract*}

\section{Introduction} \label{introduction}
Fiber Bragg gratings (FBGs) have been widely employed in sensing and telecommunication applications for several decades, and often have been a superior choice in many applications over other methods due to their compact size, high wavelength selectivity, low insertion loss, polarization insensitivity, easy integration, and simple coupling with optical fiber systems\cite{bilodeau1995all,guy1995single,eggleton1999electrically,Kok2024,Theo2024}. In astronomy, multi-channel aperiodic FBG filters, in combination with photonic lanterns (PLs), have been shown to be the promising solution to filter out the hydroxyl (OH) lines or Meinel nightglow \cite{Meinel} for ground-based spectrographs \cite{Trinh_2013,Ellis2020, Diab:21, Davenport:21, Martin, Diab_mnras, Davenport:21_2, Julian}.
 
Near-infrared (NIR) observations are important to address some of the big questions in astronomy and astrophysics, concerning dark matter and dark energy, the formation of stars, evolution of planetary systems, relationship between super-massive black-holes and galaxies, and the re-ionization of the universe \cite{ellis2008case}. However, ground-based NIR observations are significantly hindered by airglow originating from OH radicals in Earth’s atmosphere \cite{Maihara_1993}. These radicals primarily form through reactions of hydrogen and ozone from a layer about 9 km thick at an altitude of around 87 km. The rotationally and vibrationally excited OH molecules have a lifetime of a few milliseconds and vibrationally decay thus producing $4\,732$ background emission lines in the NIR. The OH lines are approximately $1\,000$ times brighter in the NIR than in visible light and fluctuate in a short time scale due to diurnal variation in temperature and density waves in the upper atmosphere \cite{ellis2008case}. To achieve low sky background observations required for estimating the velocity dispersions of dwarf galaxies, studying the gas dynamics in faint galaxies, measuring accurate redshifts, and any spectroscopic studies of faint targets \cite{dauphin2024hydroxyl}, FBG based OH filters must remove $\sim103$ OH lines in the H-band ($\SI{1.4}{\mu m}$ to $\SI{1.8}{\mu m}$) at a resolution ($\lambda/\Delta\lambda$) of $10\,000$ with high attenuation ($\approx$ \SI{30}{dB}) while minimizing loss in the interline continuum ($<$\SI{0.2}{dB}) \cite{ellis2008case}. To meet the stringent specifications of the filters, the design of those FBGs requires advanced modeling \cite{SKAAR2001,Skaar2002, Buryak:03, Bland-Hawthorn:08} and the fabrication setup should consist of high-precision stages \cite {Liu, Petermann2002, Gbadebo2018} capable of inscribing long, complex, multi-channel aperiodic gratings within the core of a single mode fiber. Moreover, replicas of these filters must be integrated with PLs so that the entire point spread function (PSF) can be accessed with the multimode fiber-end of the PLs \cite{Trinh_2013, Ellis2020, Diab:21, Davenport:21, Martin, Diab_mnras, Davenport:21_2, Julian}.
We use two complementary methods to fabricate these filters: modified elliptical Talbot interferometry, based on moving interference pattern \cite{Stepanov, Gagne:08, BuryakPatent}, and novel complex phase mask methods \cite{Rahman:20, Rahman:23,Luo}. Post fabrication, the filter's performance during operation depends on its thermal and mechanical stabilization, underscoring the need for a robust athermal packaging module. 

FBGs are inherently sensitive to temperature variation, with a typical temperature sensitivity of $\sim$ \SI{10}{pm}/$^{\circ}$C. The tolerance in thermal stability required in astronomy is dictated by the resolving power $R$ of the spectrograph used. For example, for a spectrograph with $R = 10\,000$ \cite{Hernandez}, a spectral resolution $\Delta\lambda$ of $\sim$\SI{155}{pm} at \SI{1550}{nm} will be required, i.e, the filters have to be stabilized with a tolerance $\leq$\SI{155}{pm} over the operating temperature range (typically $\sim$\SI{40}{^{\circ}C}).  The thermal stability requirement becomes further more stringent for a spectrograph with $R > 10\,000$; for example, ANDES, the high-resolution spectrograph for the Extremely Large Telescope (ELT), will operate with $R = 30\,000$ in the NIR\cite{Liske}. Therefore, FBG filters must be stabilized within $\leq$\SI{50}{pm} over the operating temperature range, so that the interline continuum is not lost. However, a much tighter tolerance than \SI{50}{pm} over the operating temperature range will be demanded when additional factors, such as the accuracy of the design and fabrication of the filters, and reproducibility, are considered.  Considering a typical temperature sensitivity of an uncompensated FBG filter of $\sim$\SI{10}{pm}/$^{\circ}$C, an athermalisation factor $F$, the ratio of the maximum wavelength shift of the compensated filter grating to the uncompensated reference grating, $F<\nicefrac{1}{8}$ must be targeted.
Athermally packaged FBGs are widely used in areas like fiber-optic sensing systems \cite{Kuang,yoffe1995passive,iwashima1997temperature,huang2003temperature}, dense-wavelength division multiplexing (DWDM) and for wavelength referencing \cite{Alxenses, Technicasa, Findlight}. However, as the FBG OH filter includes a single,  or multiple complex gratings that generate multiple aperiodic channels over a large bandwidth ($\SI{1.4}{\mu m}$--$\SI{1.8}{\mu m}$ window), the overall FBG length is considerably longer (i.e., \SI{100}{mm} -- \SI{150}{mm}) \cite{Bland-Hawthorn:08} than that typically used in sensing or telecommunication applications, necessitating a custom-designed athermal package.

We presented the design of a self-compensating athermal package in \cite{Alvarez}. The design offers flexibility in tuning the filter wavelengths, ensuring a high-precision performance of the filters. 
In this paper, we extend the work presented in \cite{Alvarez} by manufacturing and characterizing a self-temperature compensating package for the FBG filters. We studied the performance of the package over the temperature range of  \SI{-17}{^{\circ}C} to \SI{20}{^{\circ}C}, significant for ground-based astronomical use.

This paper is organized into three parts - in the first part, Section \ref{OH}, we begin by providing an observation with OH background at NIR, using examples of data from the CARMENES spectrograph. Section \ref{tempcomp} and Section \ref{model} constitute the second part, where we discuss the principle of temperature compensation in FBGs in self-compensating packages followed by the modeling of the athermal package based on finite element method (FEM). Here we also discuss the manufacturing details of the athermal package. Finally, in the third part, Section \ref{assembly}, and \ref{results_discussions}, we provide the details of the assembly of the package, experimental methods, and performance. In the performance analysis, we highlight areas for improvement, which will be addressed in the second generation of the athermal package \cite{Patent}.

\section{OH emission lines at NIR}\label{OH}

We extracted OH lines from the CARMENES spectrograph on the \SI{3.5}{m} telescope at the Calar Alto Observatory intended for earth-like exoplanet research \cite{2010SPIE.7735E..13Q}. The data was obtained from the CAHA Public Archive. The instrument has two channels: VIS-channel  (\SI{520}{nm} -- \SI{960}{nm}) and NIR-channel (\SI{960}{nm} -- \SI{1710}{nm}). We only showcase the NIR-channel data, which has the spectral resolution of $R = 80\,400$. Although the science cases that required OH filters would need much lower resolution due to the faint nature of the targets, the CARMENES resolution could offer a more accurate identification of the OH lines. Since our experiment was focused on the filters centered around 1550 nm, we intended to show the OH lines only visible in this region. However, due to a gap in the echelle orders between 1540 nm $-$ 1550 nm in the CARMENES data, we used the OH lines seen between 1553 nm and 1567 nm.

\begin{figure}[H]
    \centering
    \includegraphics[width=\linewidth]{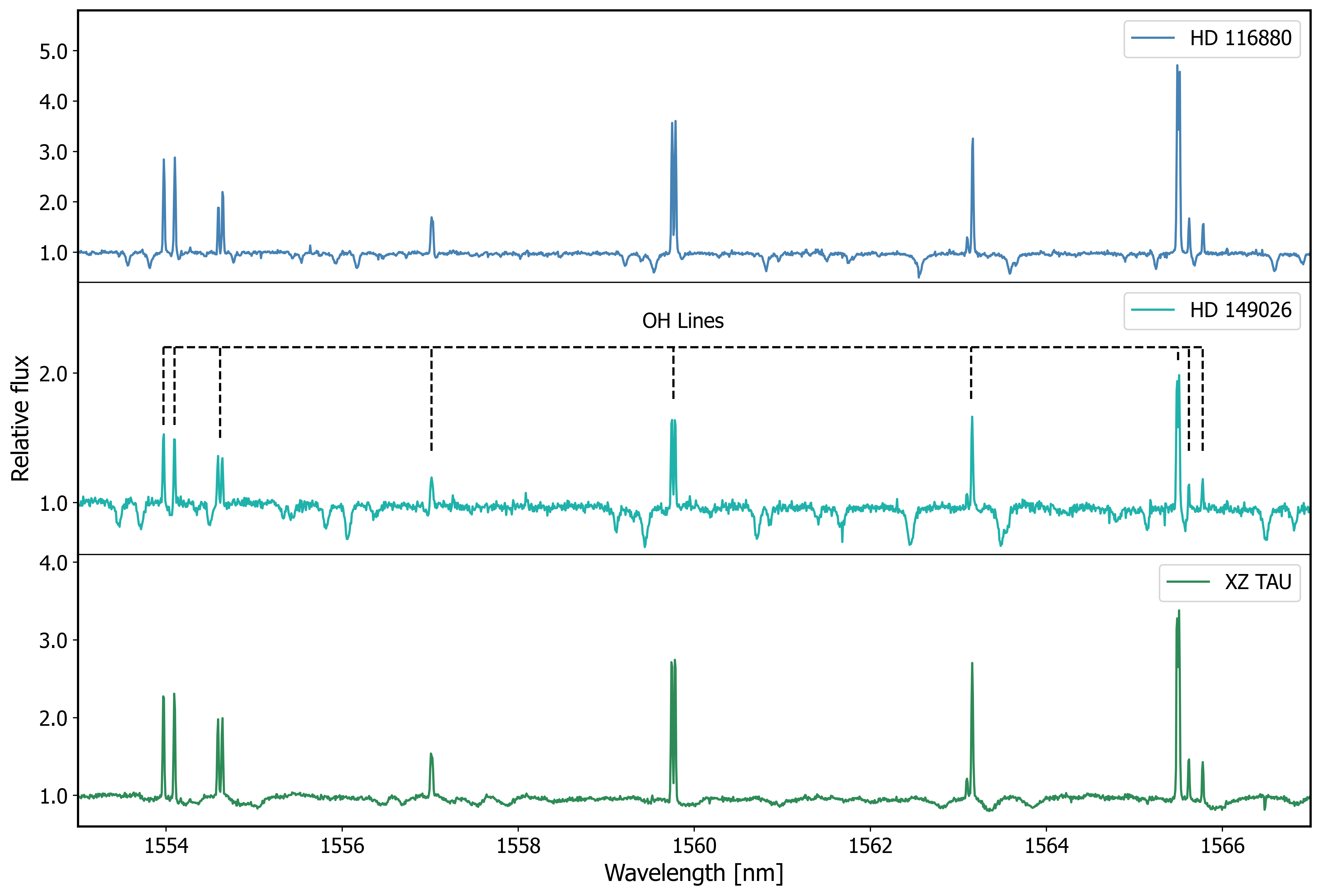}
    \caption{Identification of OH lines in three different science spectra. The OH line identification wavelengths based on \textit{Rousselot et al.}\cite{2000A&A...354.1134R}: \SI{1553.97}{nm}, \SI{1554.09}{nm}, \SI{1554.61}{nm}, \SI{1557.02}{nm}, \SI{1559.76}{nm}, \SI{1563.14}{nm}, \SI{1565.50}{nm}, \SI{1565.62}{nm}, and \SI{1565.77}{nm}.}
    \label{fig:NIR-obs}
\end{figure}

We obtained three different spectra (Fig.~\ref{fig:NIR-obs}). The first object is HD 116880, a K-type main sequence star ($H=7.92$) and a spectroscopic binary \cite{2007Obs...127..171G}, observed on the 12$^{\mathrm{th}}$ of May 2016 with the exposure time of $1\,538$ seconds. The second object is HD 149026, a G-type subgiant star ($H=6.90$) hosting a hot saturn exoplanet HD 149026b \cite{2024AJ....168..106R, 2024A&A...691A.283B}. It was observed on the 13$^{\mathrm{th}}$ of April 2019 with 248 seconds exposure time. The third object is XZ Tau, a double T-Tauri close binary system ($H=8.15$) \cite{2020A&A...638A..85R, 2021ApJ...919...55I}. T-Tauri are variable pre-main sequence stars, typically found in star forming regions. The XZ Tau data was obtained on the 21$^{\mathrm{st}}$ of January 2021 with $1\,793$ seconds of exposure. The OH lines were identified using the line list from \textit{Rousselot et al.}\cite{2000A&A...354.1134R},  observed at a resolution of $R=8\,000$. The spectra were normalized to their pseudo-continuum level where no emission, nor absorption lines are present, as shown in Fig.~\ref{fig:NIR-obs}.

We found that CARMENES could resolve some of the lines that were actually doublets. However, that would not be important when considering the future science case with $R \sim 10\,000$, closer to the spectrograph resolution employed for the line list. We could also show that the OH lines have a very consistent wavelength between three different object taken at different times. On the other hand, the intensity of the OH lines varied at different times. If one considers a low-resolution spectrograph and the line-broadening effects of astrophysical instruments, a post-observation OH-line removal would be particularly challenging, especially when there are important science lines in the vicinity. Meanwhile, the OH lines could not be thoroughly removed even from high dispersion spectroscopy, since the OH emissions are much brighter compared to the interline continuum. Even a small amount of scattered OH light will hinder the interline continuum \cite{ellis2008case}. Note that our example spectra were taken from really bright nearby stars, while the demand for OH filters mainly comes from much fainter objects. Therefore, removing the OH lines using the FBGs before the light enters the spectrographs provides a promising solution to the problem.

\section{Temperature compensation in FBGs} \label{tempcomp}

The FBGs are formed by photoinduced periodic or aperiodic refractive index modulation within the fiber core of a single-mode fiber. When the Bragg condition $\lambda_\text{B} = 2n_\text{eff}\Lambda$ is satisfied, the grating reflects light at the Bragg wavelength $\lambda_\text{B}$, where $n_\text{eff}$ is the effective refractive index and $\Lambda$ is the period of the grating. Therefore, any perturbation on $n_\text{eff}$ or $\Lambda$ results in a shift in the Bragg wavelength $\lambda_\text{B}$. The mechanical strain or temperature variation gives rise to a change in both the effective refractive index and the period of the FBG, leading to a Bragg wavelength shift $\Delta \lambda_\text{B}$, which is given by \cite{huang2003temperature}:

\begin{equation}\label{eq:1}
    \frac{\Delta \lambda_\text{B}}{\lambda_\text{B}} = (1-p_\text{e})\epsilon+(\alpha+\xi)\Delta T
\end{equation}

where $p_\text{e}$ is the photoelastic constant of the optical fiber, $\epsilon$ is the strain induced on the fiber, $\alpha$ is the thermal expansion coefficient of the fiber, $\xi$ is the thermo-optic coefficient of the fiber, and $\Delta T$ is the temperature variation, respectively.

Temperature compensation in an FBG is achieved when the Bragg wavelength shift $\Delta \lambda_\text{B}$ caused by temperature variations is counteracted by an applied strain that induces an opposing shift in $\lambda_\text{B}$. The compensating strain can be introduced to the FBG using a packaging system with materials of dissimilar coefficients of thermal expansion (CTE) or a thermoelectric material and a feedback system to stabilize the temperature of the FBGs. The first approach is preferred for many astronomical applications due to its higher reliability, lower complexity, absence of power dependency, minimal maintenance, and cost-effectiveness. 
In our design, we employ a prestretched FBG in a double-end fixed packaging arrangement\cite{Alvarez} chosen for its simplicity, straightforward operation, and, most importantly, for its ability to eliminate chirp along the length of the FBG. The behavior of an FBG in such an athermal unit is described in \cite{Lachance}, and is defined by the following equations: Eq.~(\ref{eq:2}) - Eq.~(\ref{eq:5}):
\begin{equation}\label{eq:2}
\frac{\Delta\lambda_\text{B}}{\lambda_\text{B}}= \xi\Delta{T}+\alpha_\text{a}\Delta{}T-p_\text{e}(\alpha_\text{a}-\alpha)\Delta{}T,
\end{equation}
where, $\alpha_\text{a}$ represents the CTE of the athermal package, characterizing the thermal behavior of the distance between the glued points of the fiber inside the athermal package.
Athermalization is reached when the right-hand side of Eq.~(\ref{eq:2}) is equal to $0$, for example:

\begin{equation}\label{eq:3}
\alpha_\text{a}=-\frac{\alpha p_\text{e}+\xi}{1-p_\text{e}}
\end{equation}

In general, in a multimaterial athermal structure, where athermalisation is achieved by a combination of $N$ number of elements, the distance between the glued points of the fiber inside the structure is given by Eq.~(\ref{eq:4}):
\begin{equation}\label{eq:4}
{\text{$L$}_\text{a}} = \sum_{i=1}^{N}\
{c_i}{L_i},
\end{equation}
where, ${c_i}={+1}$ or ${-1}$ defined by the geometry of the structure, and ${L_i}$ is the length of the ${i^{th}}$ element of the structure. Finally, the CTE of the package, $\alpha_\text{a}$ is given by Eq.~(\ref{eq:5}):

\begin{equation}\label{eq:5}
{\alpha_\text{a}} = \frac{1}{\text{$L$}_\text{a}}\frac{d\text{$L$}}{dT}= \frac{\sum_{i=1}^{N}\
{c_i}{\alpha_{i}}{L_i}}{\sum_{i=1}^{N}\
{c_i}{L_i}},
\end{equation}
where, $\alpha_\text{i}$ is the CTE of the ${i^{th}}$ material used in the athermalization structure. In this work, we manufactured a dual-material hollow tube with materials of dissimilar CTE, where the FBG filter is bonded at both ends of the tubular structure as designed in \cite{Alvarez}. In our design, Eq.~(\ref{eq:5}) is then reduced to two materials and their corresponding lengths. The next sections provide the modeling and manufacturing details of the athermal package.


\section{Modeling and manufacturing of the athermal package}\label{model}
\subsection{Modeling of the athermal package}
 We modeled the athermal package using the finite element method (FEM) to find the appropriate combination of length and material of the various components of the athermal package needed to compensate for the effect of temperature variations in the FBG. The FEM analysis was based on the experimentally determined strain and temperature sensitivity values of FBG \cite{Alvarez}. 
\begin{figure}[H]
\centering\includegraphics[width=0.9\textwidth]{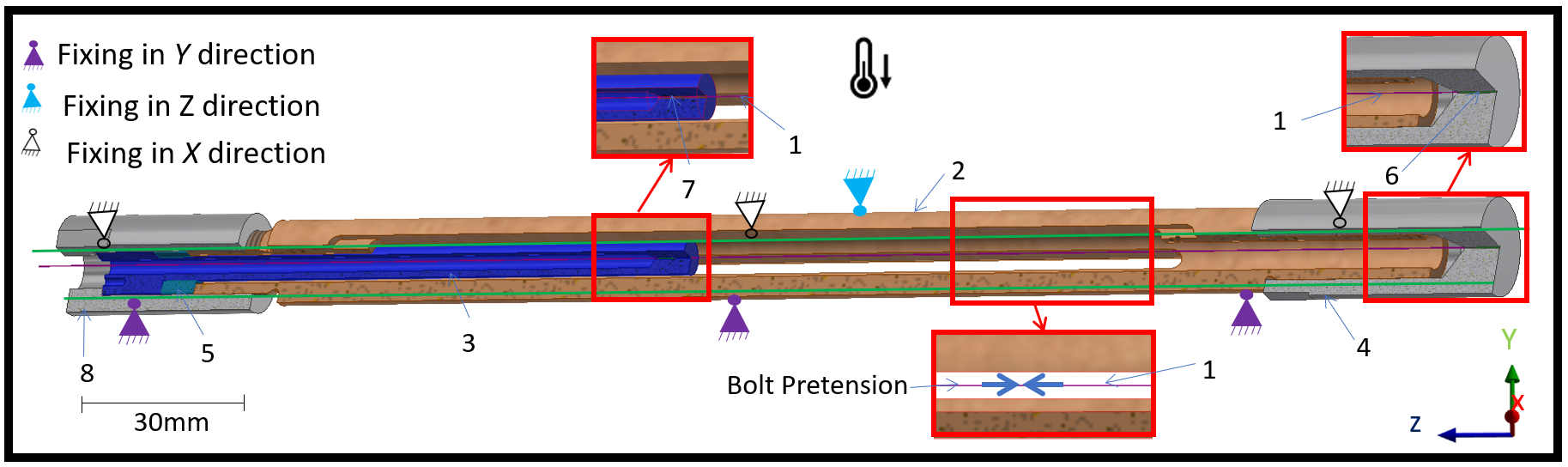}
\caption{Boundary conditions for the FEM analysis. The different components/geometries composing the FEM model: Optical fiber (1), tensioner (2), angle adjuster (3), fiber holder (4), precision shims (5), glue on fiber holder (6), glue on angle adjuster (7), and safety cap (8).}
\label{fig:fem1}
\end{figure}
Using a finite element analysis (FEA) software, we first programmed different parameters as conditions, such as the length and CTE of the different components, and then studied the effects of changes in these parameters on the overall model. 
\subsubsection{FEM model} 

Figure~\ref{fig:fem1} shows the different mechanical components of the model. Most of the geometries or components in the FEM model were connected by a ``bond'' contact condition, except for the conditions, between the angle adjuster and precision shim components, angle adjuster 3 and safety cap 8 components, and the precision shim and tensioner components, where the ``no separation'' contact condition was used. 
In FEA, ``bond'' contact and ``no separation'' contacts define how two surfaces interact during a simulation. In a ``bond'' contact, neither sliding nor separation is allowed under any loading conditions. In a ``no separation'' contact, the surfaces remain in contact but are allowed to slide freely against each other without friction.
Hexahedral and tetrahedral elements were used to mesh the FEM model. It is essential to constrain the model by restricting all six degrees of freedom (as illustrated in Fig.~\ref{fig:fem1}).
For parts and components bigger than \SI{5}{mm}, the mesh element size is \SI{1}{mm}, and for the remaining parts, such as the shims, optical fiber, and glue, the mesh element size is \SI{0,1}{mm}. 
There were five boundary conditions as shown in Fig.~\ref{fig:fem1}, which include a) three displacement constraints  that restrict  all parts of the model in six degrees of freedom of space, b) a bolt preload applied along the optical axis of the fiber, and c) an external thermal condition that affects  all parts of the model. 
The bolt pretension ($\SI{0.125}{N}$) provides the necessary pretension acting on the fiber, as the fiber has to be mounted with a certain preload to ensure that the tension is released when the temperature increases and the fiber can relax. The FEM simulation is done in two sequential steps: first, the fiber is loaded with the bolt pretension, and then a temperature change from $20^{\circ}$C to $-20^{\circ}$C ($|\Delta T|= \SI{40}{^{\circ}C}$) is applied. In the simulation, we considered the temperature change from $20^{\circ}$C to $-20^{\circ}$C, as this range covers the temperature variations for ground-based astronomical applications.
\begin{figure}[H]
\centering\includegraphics[width=0.8\textwidth]{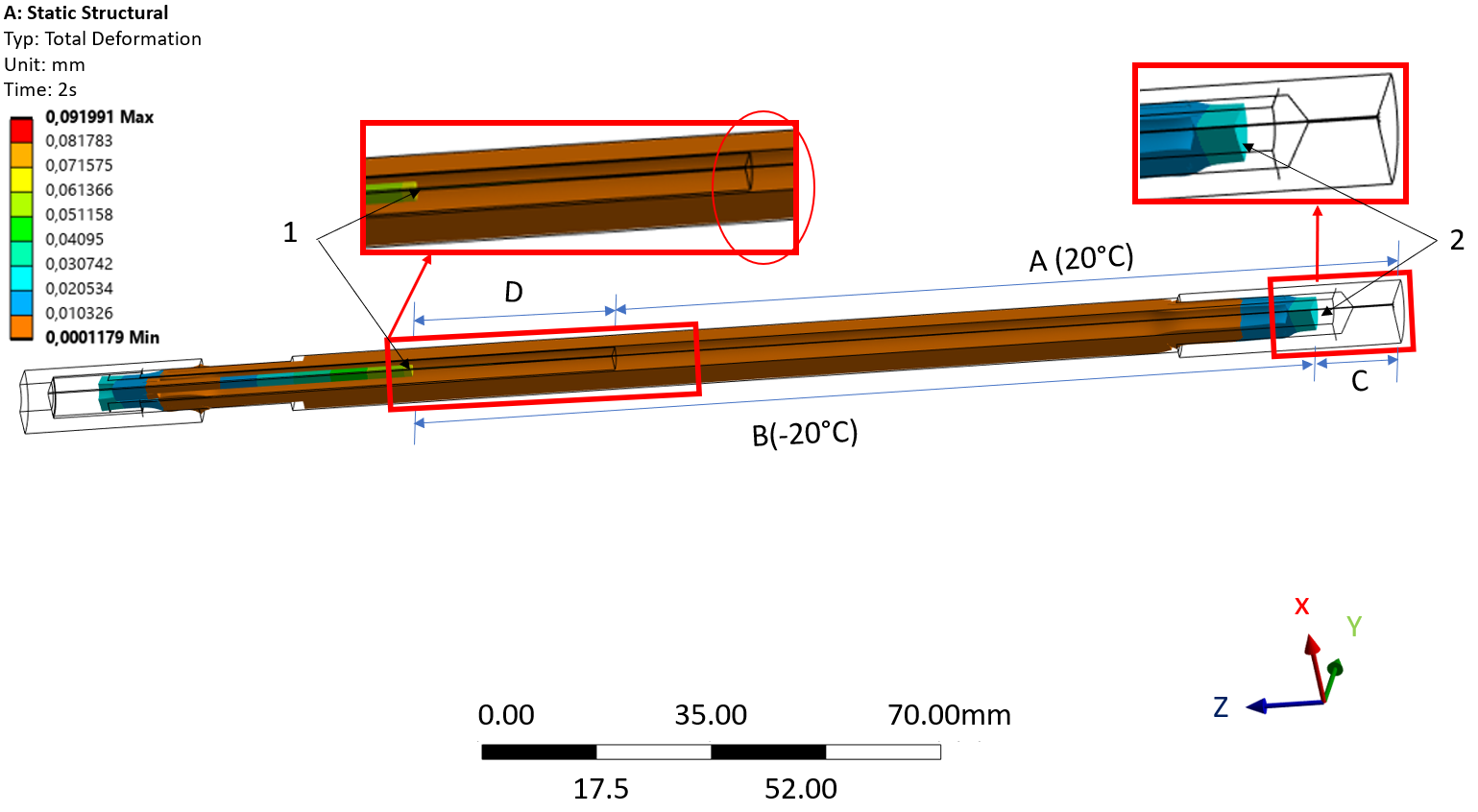}
\caption{The mechanical structure of the athermal package deforms when the boundary conditions are applied to them in the FEM analysis. The frame represents the parts at room temperature. The colors indicate the amount of deformation experienced by the parts. The values on the deformation color scale are given in \SI{}{mm} (Deformation image magnified 500 times).}
\label{fig:femresults}
\end{figure}
\subsubsection{FEM results}
Figure~\ref{fig:femresults} highlights four lengths, labeled as A, B, C, and D that are essential for the athermal package. Length A represents the distance between the two glued points where the optical fiber is bonded. Length A measures $\SI{118.37}{mm}$ at an ambient temperature $20^{\circ}$C. Length B corresponds to the separation between surfaces 1 and 2 (as marked in Fig.~\ref{fig:femresults}) after the temperature changed from $20$ $^{\circ}$C to $-20$ $^{\circ}$C. B measures $\SI{118.41}{mm}$ at $-20$ $^{\circ}$C. Lengths C and D refer to the fiber holder and angle adjuster, respectively. All components undergo dimensional changes in response to temperature change ($|\Delta T|=\SI{40}{^{\circ}C}$). Our FEM calculations indicate that A ($\SI{118.37}{mm}$) increases to B ($\SI{118.41}{mm}$) with a temperature drop of $40$ $^{\circ}$C, corresponding to a $\SI{1}{\mu m}$ increase in length per $\SI{-1}{^{\circ}C}$ change. In contrast, with decreasing temperature, lengths C and D contract by $\SI{0.026}{mm}$ and $\SI{0.064}{mm}$, respectively. As lengths C and D shrink, the resulting expansion of length A to B causes the optical fiber to stretch, inducing strain on the FBG. This strain counteracts the wavelength shift of the FBG caused by the temperature drop and is the key to achieving athermalization. The opposite effect is observed when the temperature increases.


This selective distance change between the two surfaces, $1$ and $2$ (Fig.~\ref{fig:femresults}), after the temperature change, which we refer to as athermalization, can be achieved with different combinations of component lengths and materials. It is important to note that materials with negative CTE can be explored; however, we have chosen conventional materials with positive and dissimilar CTEs primarily due to their ease of machining, repeatability, and cost-effectiveness. The combinations chosen for the prototype are summarized in Table~\ref{tab:finalconfig}. All lengths were calculated at $20^{\circ}$C, and the glue was modeled with the same physical properties as the fiber due to its small quantity. 

\begin{table}[htbp]
\centering
\caption{Materials and lengths selected for the fiber displacement of \SI{0,001}{mm/^{\circ}C}.} 
\label{tab:finalconfig}
\resizebox{\textwidth}{!}{
\begin{tabular}{llclc}
\hline
 Nr&\multicolumn{1}{l}{Component}                                                                      & Calculated length  [mm]& \multicolumn{1}{c}{Material} & \ CTE [$10^{-6}$/ $^{\circ}$C]\\
 \hline
 1& Optical fiber (between the glue points)              &  118.373         & Glass SiO$_2$                      & 44.18\\
 2&  Tensioner                  &  186.500         & Invar\textsuperscript{\textregistered}~36                              &  1.2\\
 3& Angle adjuster              &   89.147         & Aluminium (AlSiMgMn)               & 23.4\\
 4& Fiber holder                &   33.150         & Aluminium (AlSiMgMn)               & 23.4\\
 5& Precision shims             &    5.020         & Aluminium (AlSiMgMn)               & 23.4\\
 6& Glue (applied to two spots)                   &    3.500         & assumed to be Glass SiO$_2$        & 44.18\\
 7& Safety cap                  &   26.150         & Aluminium (AlSiMgMn)               & 23.4 \\
 \hline
 \end{tabular}
}
\end{table}
\subsection{Manufacturing of the athermal package} 
All, except for one, components of the athermal package are made of aluminum (AlSiMgMn EN AW-5019, material number 3.2315)\cite{proprieties_alu}. Only the component, tensioner is made of Invar\textsuperscript{\textregistered}~36. For the material of the optical fiber, fused silica is assumed. Further information on the length of the components used is provided in Table~\ref{tab:komp_messung1}.

\begin{table}[htbp]
\begin{longtable}{p{2.2cm}p{2cm}p{2.5cm}p{4.5cm}}

\caption
{Different components of the athermal package and their respective dimensions.}
\label{tab:komp_messung1}\\
\hline
Component name& Image& Measured length& Measuring instrument\\
\hline
\hline
Tensioner& \begin{tabular}[c]{@{}c@{}}\\  \includegraphics[width=0.13\textwidth]{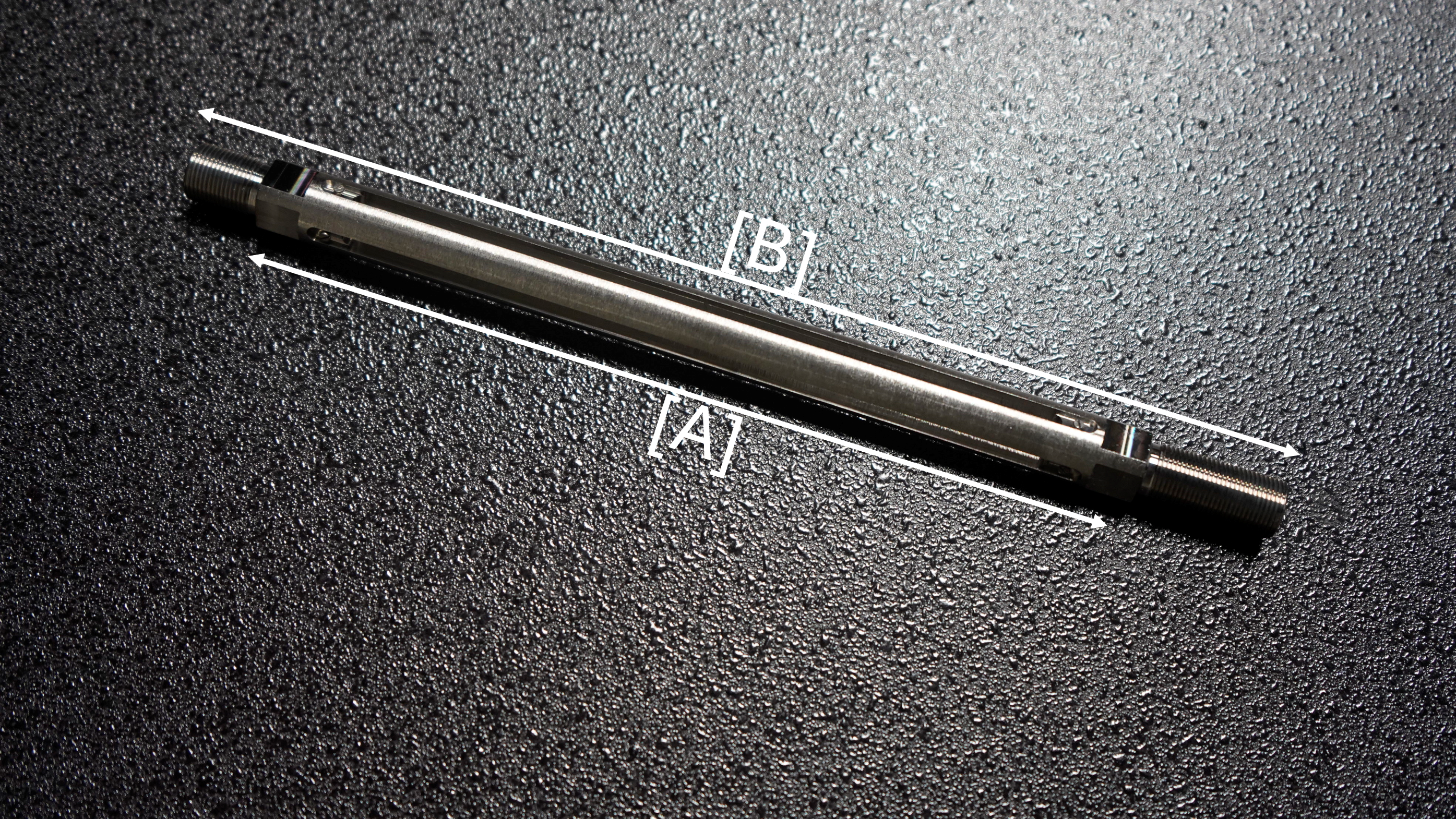}\end{tabular} & [A]$\SI{163.50}{mm}$ [B]$\SI{186.51}{mm}$ &[A] digital caliper\newline [B] digital micrometer Mitutoyo\\
Fiber holder& \begin{tabular}[c]{@{}c@{}}  \includegraphics[width=0.13\textwidth]{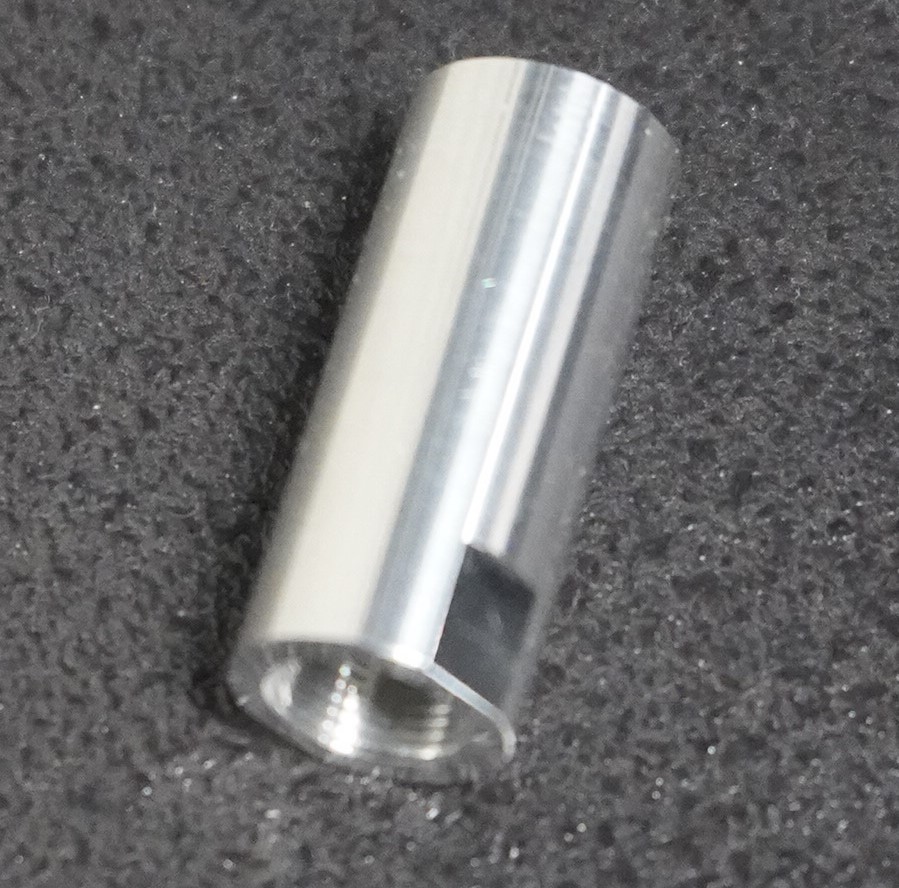}\end{tabular}& $\SI{33.50}{mm}$&digital micrometer Mitutoyo\\
Precision shims& \begin{tabular}[c]{@{}c@{}}  \includegraphics[width=0.13\textwidth]{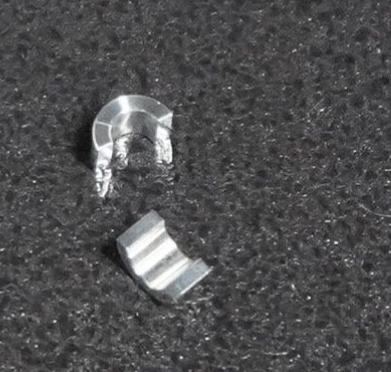}\end{tabular}& not measured & N/A\\
Safety cap& \begin{tabular}[c]{@{}c@{}}  \includegraphics[width=0.13\textwidth]{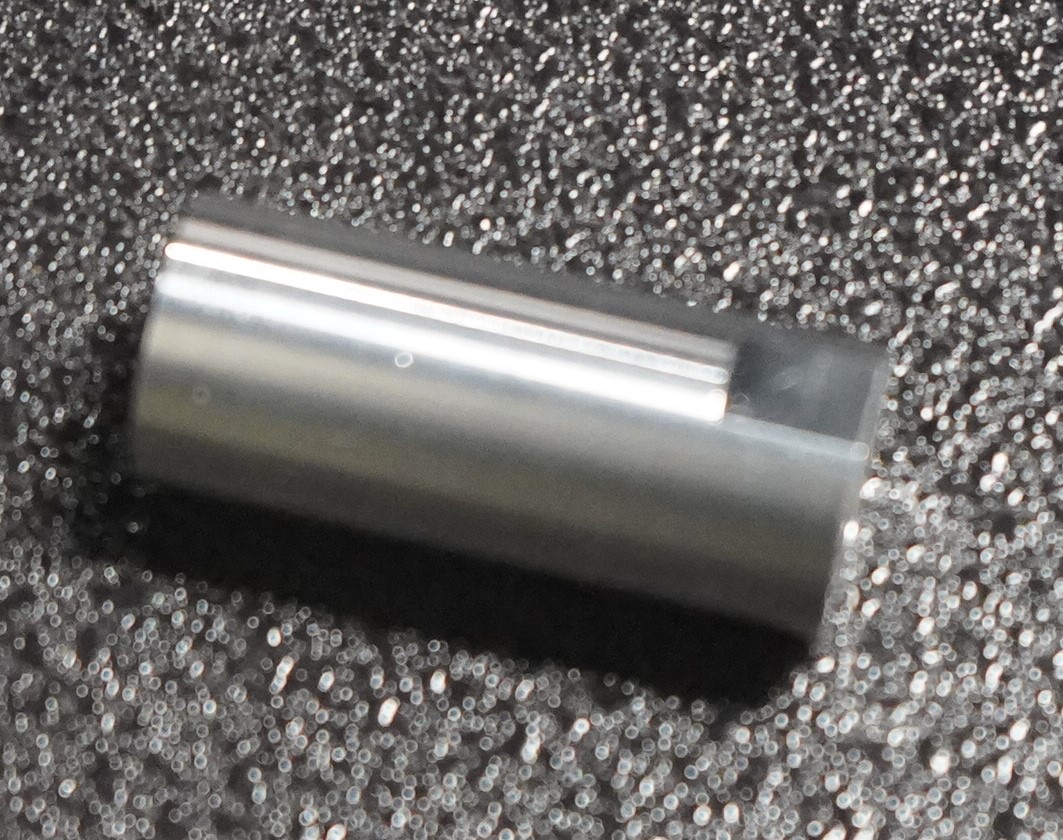}\end{tabular}& $\SI{28,00}{mm}$& digital micrometer Mitutoyo\\
Angle adjuster& \begin{tabular}[c]{@{}c@{}}  \includegraphics[width=0.13\textwidth]{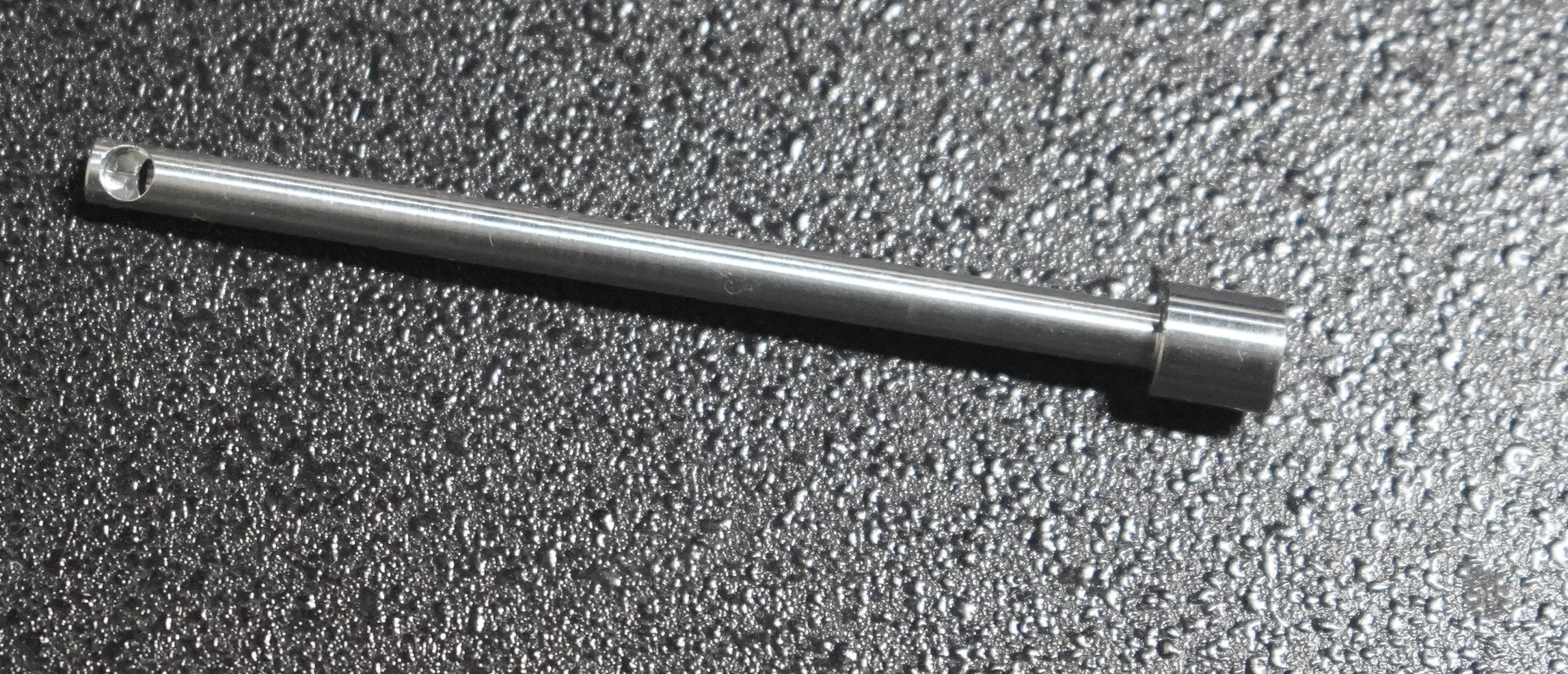}\end{tabular}& $\SI{89,162}{mm}$& digital micrometer Mitutoyo\\
\raggedright Precision shims in nominal lengths & \begin{tabular}[c]{@{}c@{}}  \includegraphics[width=0.13\textwidth]{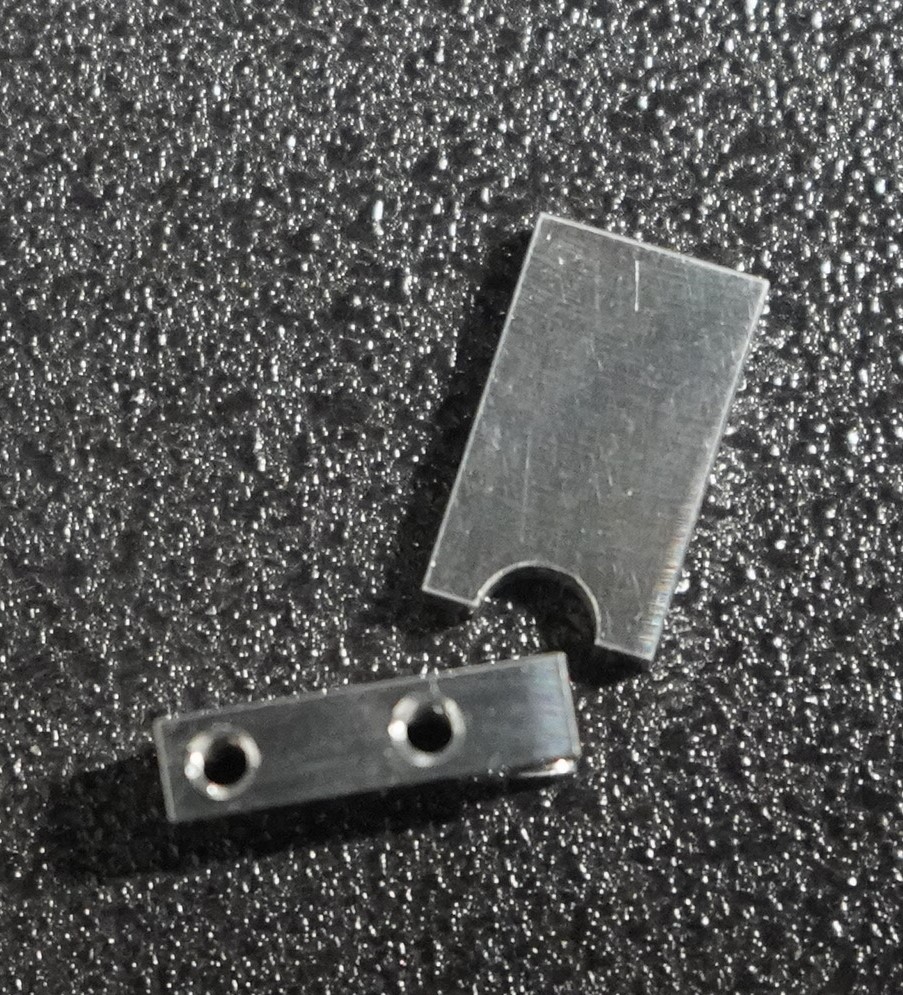}\end{tabular}& $\SI{4,98}{mm}$ & digital micrometer Mitutoyo\\
\hline
\end{longtable}
\end{table}

The FEM simulation yielded an optical fiber displacement of $\SI{1}{\mu m/^{\circ}C}$, while the experimentally measured value was $\SI{1.041}{\mu m/^{\circ}C}$, which was an approximate difference of $4\%$. Material imperfections and the length tolerances required during manufacturing play an important role in the final test results. In the FEM model, exact dimensions and perfect materials are assumed, which may not fully reflect real-world conditions. Additionally, the accuracy of the various measuring instruments used can influence the overall outcome \cite{Alvarez}. However, this deviation has not been considered relevant for the manufacturing of the prototype because, in reality, the length of the parts depends on the manufacturing and measurement tolerances (which are in the micrometer range), as well as many other factors that affect thermal compensation. These include, for example, the uniformity of the glue layer or the concentricity of the optical fiber in the bonding hole after the application of the preload and the curing of the glue.

\label{subsec:manufacturingofathermalpackage}
\begin{figure}[htbp]
\centering\includegraphics[width=0.6\textwidth]{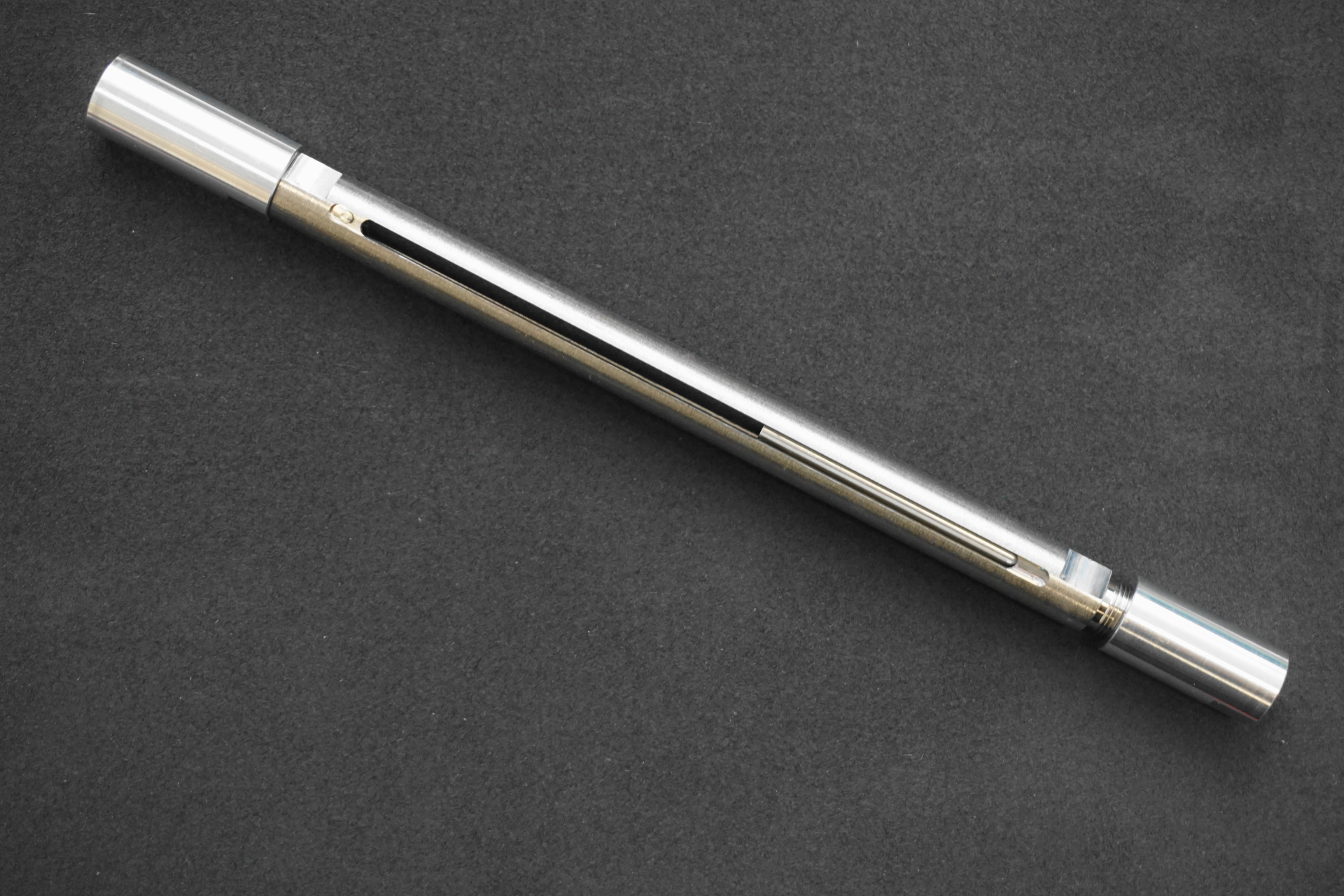}
\caption{Prototype manufactured and assembled at the Leibniz Institute for Astrophysics Potsdam (AIP).}
\label{fig:fristathermalpackage}
\end{figure}


The various components of the athermal package were manufactured by the workshop of the Leibniz Institute for Astrophysics Potsdam (AIP). The workshop has a temperature controlled environment (with $\sim$ $20^{\circ}$C constant room temperature) and is specialized in precision mechanics. The components can be manufactured and measured with an accuracy of almost a hundredth of a millimeter. After manufacturing, the components were measured longitudinally in a temperature-controlled room. The results of the measurement and the measuring instruments used are summarized in Table~\ref{tab:komp_messung1}. Figure~\ref{fig:fristathermalpackage} shows the assembled prototype.

\section{Assembly and experimental method }\label{assembly}
\subsection{Inscription of FBGs and preparation of the package}
\begin{figure}[H]
\centering\includegraphics[width=12cm, clip]{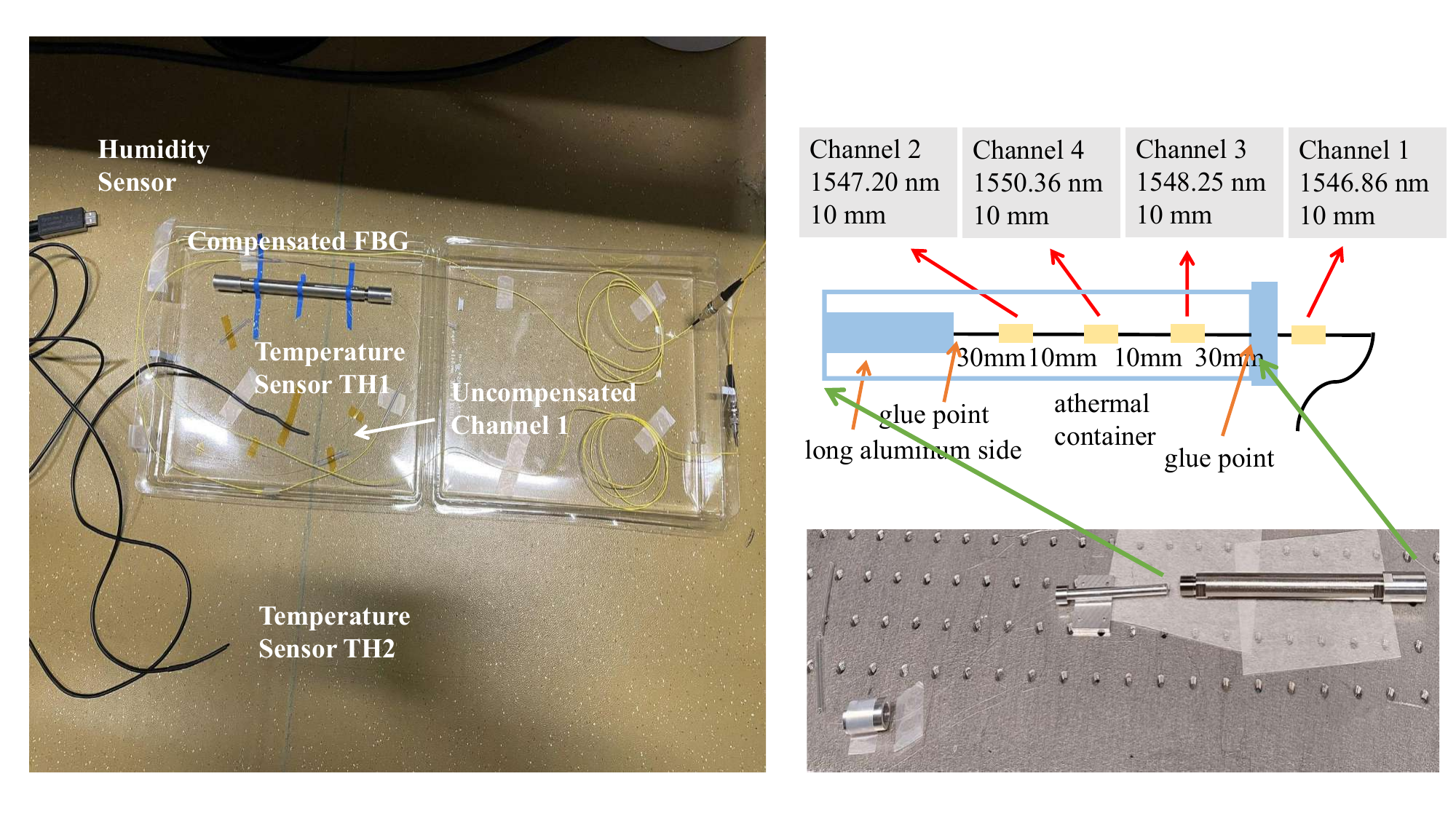}
\caption{Arrangement of the FBGs within the athermal package, and positions of the temperature sensors TH1 and TH2, and the humidity sensor.}
\label{fig:athermal_pack_layout}
\end{figure}

In this study, the FBGs were fabricated in a boron-doped photosensitive fiber using the phase mask technique, utilizing a continuous wave ArI UV laser (Coherent Sabre FRED) \cite{laser_coherent} operating at \SI{244}{nm} \cite{hill1993bragg}. We inscribed four FBG filters, labeled as Channel $1$, $2$, $3$, and $4$ in the photosensitive fiber. Each channel occupies around \SI{10}{mm} along the fiber, and the Bragg wavelengths for Channels $1$, $2$, $3$, and $4$ were \SI{1546.86}{nm}, \SI{1547.20}{nm}, \SI{1548.25}{nm}, and \SI{1550.36}{nm}, respectively. Figure~\ref{fig:athermal_pack_layout} shows how the FBGs were assembled to the athermal package. 

\begin{figure}[h!]
\centering\includegraphics[width=12cm]{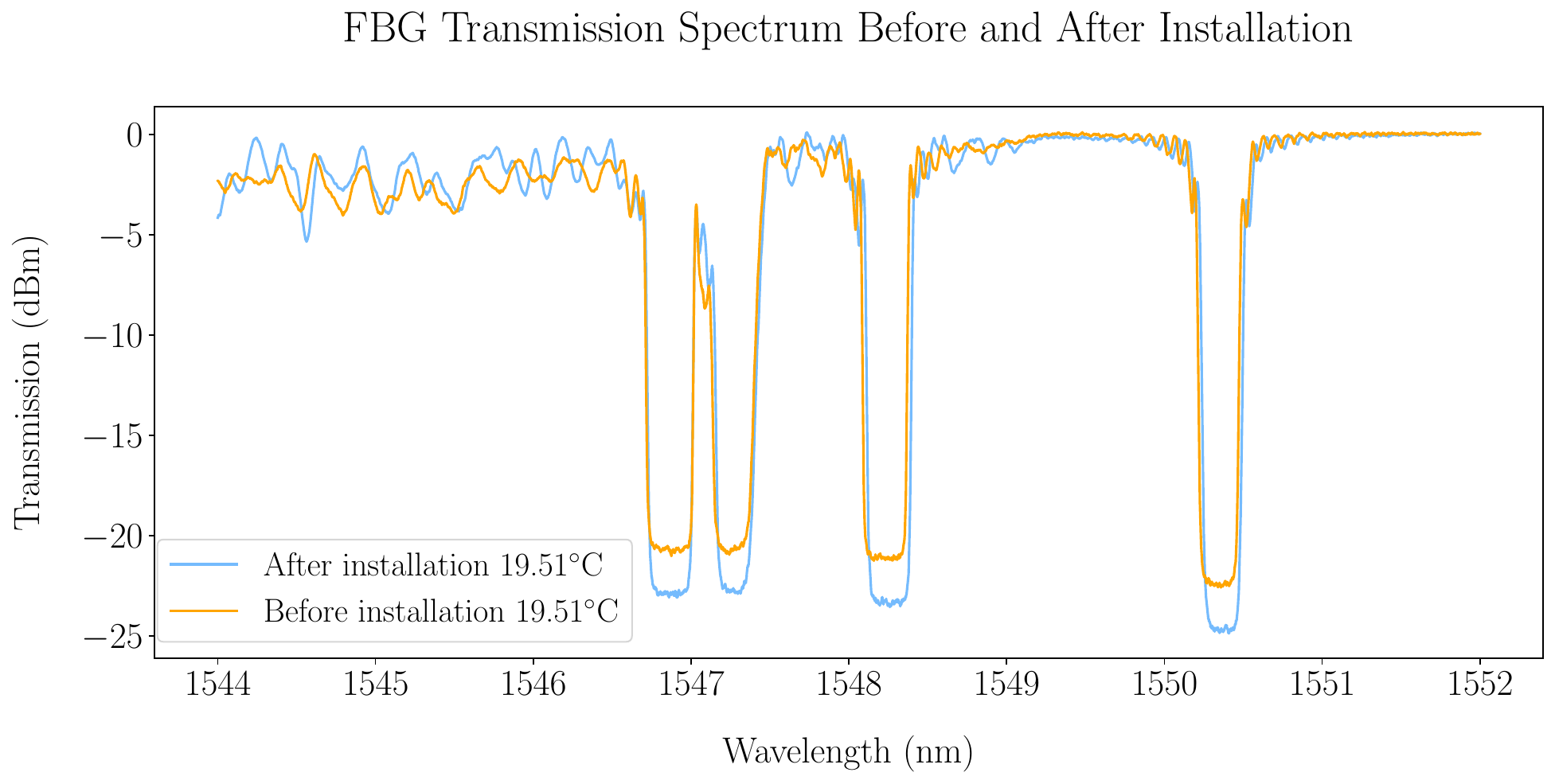}
\caption{Transmission spectrum of the FBG at room temperature \SI{19.51}{^{\circ}C} before and after the FBG was bonded to the athermal package with cyanoacrylate glue.}
\label{fig:FBG_transmission_release_spec}
\end{figure}

Channel $1$ was positioned outside the athermal package and served as a reference temperature sensor, as its center wavelength responds linearly to temperature changes. The remaining FBG channels ($2$, $3$, and $4$) were placed inside the athermal package. Channel $2$, located opposite Channel $1$, was bonded to the long aluminum rod using cyanoacrylate glue, while Channel $4$ is positioned between Channel $2$ and Channel $3$. Since this athermal package was designed for long-length, multi-channel FBGs, it is essential to evaluate its temperature compensation at different points along the fiber bonded within the package. To address this point, Channel $2$, $3$, and $4$ were evenly distributed throughout the athermal package. Figure~\ref{fig:FBG_transmission_release_spec} shows the transmission spectra of the FBG channels at room temperature $19.51$$^{\circ}$C before and after the FBG was attached to the athermal package via cyanoacrylate glue. We observed non-uniform shifts among Channel $2$, $3$, and $4$ before and after the bonding of the FBGs to the athermal package, which we will address in Section \ref{Analysis}.

\begin{figure}[t!]
\centering\includegraphics[width=12cm]{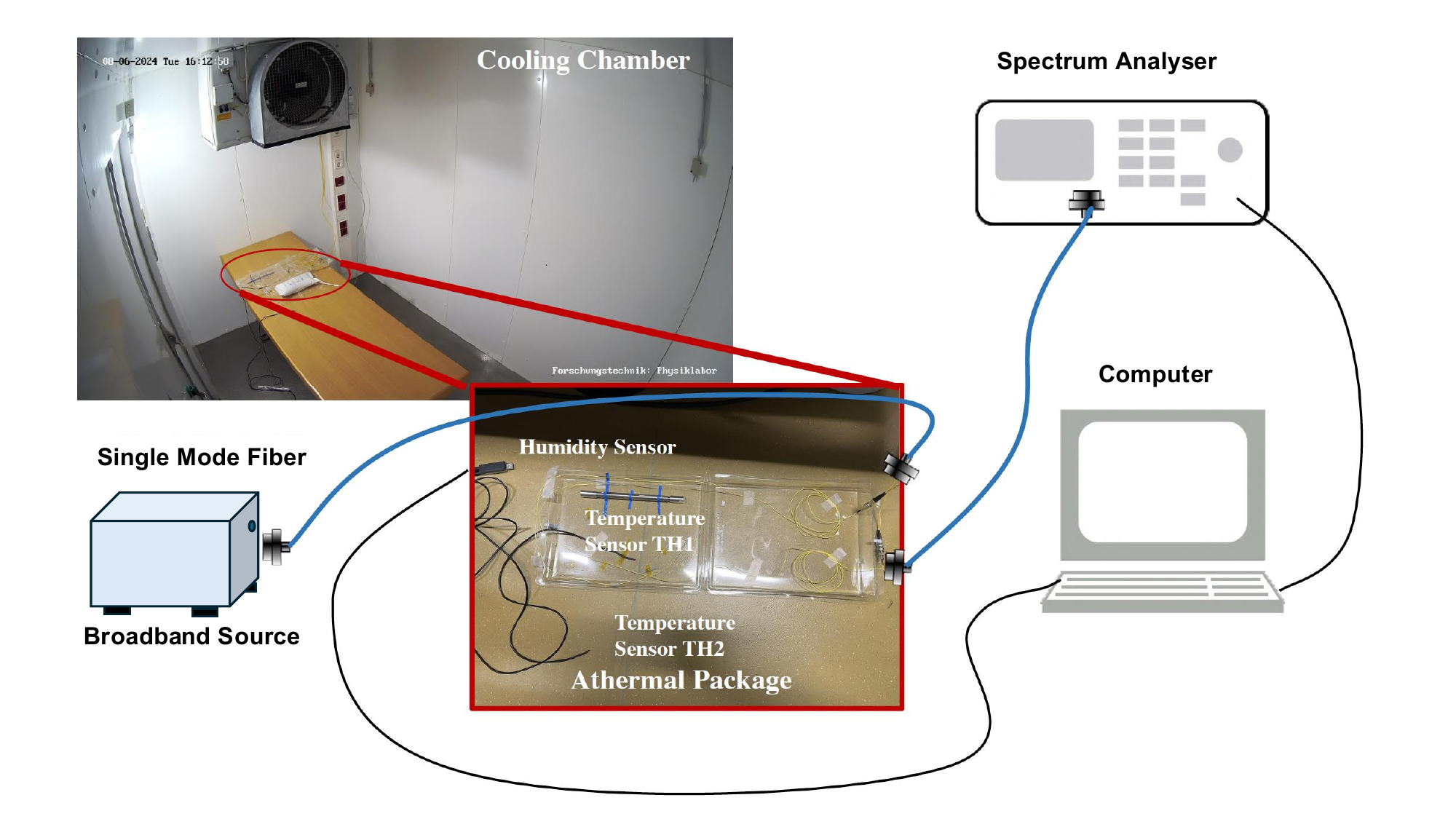}
\caption{Temperature compensation experimental setup for measuring the transmission spectrum of the FBG.}
\label{fig:FBG_connection}
\end{figure}

The athermal package and reference Channel $1$ were placed inside a plastic box where an electronic temperature sensor, TH1 probe (Thorlabs TSP01), was positioned near the uncompensated Channel $1$ (as shown in Fig.~\ref{fig:athermal_pack_layout}). This plastic box assembly was then placed in the cooling chamber 
(manufacturer: Dresdner K\"u{}hlanlagenbau GmbH
). Since humidity was not controlled in this chamber, the plastic box was wrapped with a plastic film to minimize humidity fluctuations and prevent ice formation during temperature changes. Additionally, silica gel packets were placed inside the box to absorb moisture. The athermal package was placed on the table inside the cooling chamber with a second temperature sensor probe TH2 (Thorlabs TSP01) placed close to it. A humidity sensor, (Thorlabs TSP01) was used to measure the humidity inside the cooling chamber. The FBGs were connected to a broadband source (Thorlabs ASE730) and an optical spectrum analyzer (OSA) (Apex AP2683A) with a resolution of $\SI{20}{pm}$. The temperature sensors TH1 and TH2, the humidity sensor, and the OSA were connected to a computer for data acquisition. This setup is illustrated in Fig.~\ref{fig:FBG_connection}.  

\subsection{Experiment: cooling and heating cycles}
The temperature stability of the FBG inside the athermal package was tested in three cycles inside the cooling chamber. The cooling chamber varied temperature from room temperature ($\sim$ \SI{20} {^{\circ}C}) down to \SI{-20} {^{\circ}C}, with an uncertainty of $\pm$ \SI{0.01}{^{\circ}C}. The cooling chamber was not equipped with a humidity controller, making moisture a likely factor affecting the measurements. While the temperature could be reduced in controlled steps, the rate of temperature change was not regulated. The subsequent heating process relied on natural warming, and after each cooling cycle, accumulated moisture was manually removed using absorbent cloths. Three cooling cycles were conducted, with the complete cooling and heating process spanning $21$ days (Cycle $1$ spanned more days compared to Cycle $2$ and $3$, as explained next).

\subsubsection{Cycle $1$}
In Cycle $1$, the cooling process started at room temperature (\SI{19.43}{^{\circ}C}), and we reduced the temperature by a cooling step of \SI{5}{^{\circ}C}. After each \SI{5}{^{\circ}C} temperature decrement, we allowed a full day for thermal stabilization. Transmission spectrum and temperature data were recorded over time, with synchronized timestamps for both the FBG transmission spectra and temperature measurements, confirming perfect data alignment between the two systems (i.e., FBG spectra recording set up and temperature recording set up). The temperature was measured every second and was averaged for each spectrum, as the OSA required \SI{6}{s} for sweep and data recording. We repeated this process until the temperature reached \SI{-20}{^{\circ}C} inside the cooling chamber. Despite the temperature of the chamber being set to \SI{-20} {^{\circ}C}, the internal temperature of the athermal package stabilized at approximately \SI{-16}{^{\circ}C}. After cooling, a natural heating process was allowed to happen in the cooling chamber, as it lacked an active heating mechanism. 
 
\subsubsection{Cycle $2$}
In Cycle $2$ and Cycle $3$, unlike Cycle $1$, the temperature was decreased continuously, without allowing it to stabilize at every \SI{5}{^{\circ}C}.
However, we observed a non-uniform shift in the wavelengths of Channel $2$, $3$, and $4$ in Cycle $1$ between \SI{20}{^{\circ}C} and \SI{15}{^{\circ}C} during the temperature cooling process. To validate this observation in Cycle $2$, during the cooling process from the room temperature, we first allowed the temperature to stabilize at  \SI{15}{^{\circ}C}, before further cooling to \SI{-16.55}{^{\circ}C}. After reaching \SI{-16.55}{^{\circ}C}, a natural heating process began in the cooling chamber. 

\subsubsection{Cycle $3$}
In Cycle $3$, the temperature was lowered directly from room temperature (\SI{19.77}{^{\circ}C}) to \SI{-16.95}{^{\circ}C}, without any intermediate stabilization during its cooling process. Then the cooling chamber was allowed to naturally warm back up to room temperature. 

As compared to Cycle $1$, Cycle $2$ has only three temperature stabilized data at \SI{20.09}{^{\circ}C}, \SI{15.13}{^{\circ}C}, and \SI{-16.55}{^{\circ}C}, while Cycle $3$ has two stabilized data at \SI{19.88}{^{\circ}C} and \SI{-16.99}{^{\circ}C}.
The relative humidity (RH) data was only recorded since the cooling chamber could not actively control humidity levels. 
In the next section we present the results and discuss the compensation performance of the athermal package, while comparing 
the channels with their respective physical position inside the athermal package (Fig.~\ref{fig:athermal_pack_layout}).

\begin{figure}[t!]
\centering

\begin{subfigure}{.49\textwidth}
    \centering
    \includegraphics[width=\linewidth]{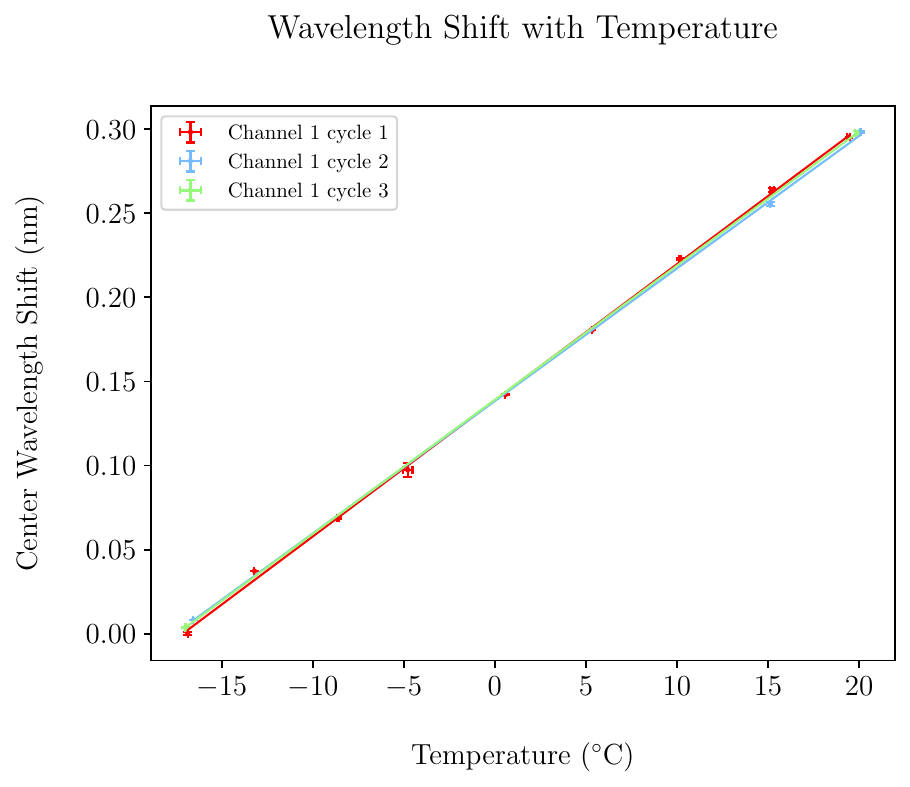}
    \captionsetup{justification=centering, singlelinecheck=false, position=bottom}
    \caption*{\makebox[1.15\linewidth][c]{(a)}}
\end{subfigure}%
\begin{subfigure}{.5\textwidth}
    \centering
    \includegraphics[width=\linewidth]{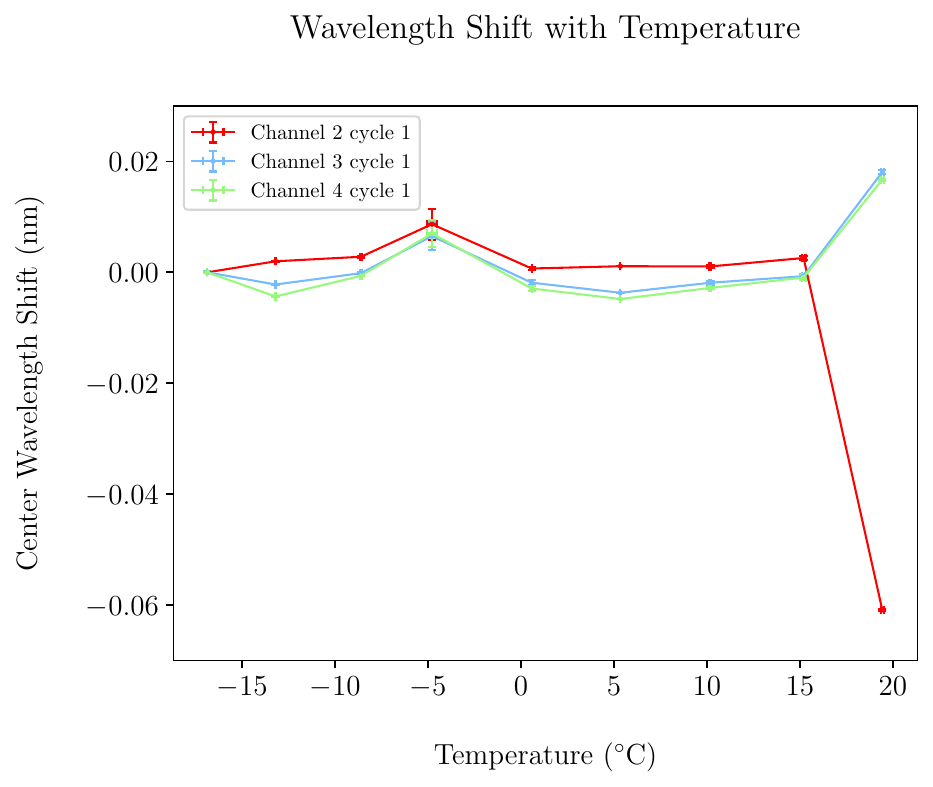}
    \captionsetup{justification=centering, singlelinecheck=false, position=bottom}
    \caption*{\makebox[1.15\linewidth][c]{(b)}}
\end{subfigure}

\caption{Shift in Bragg wavelengths at stabilized temperatures,  recorded at every \SI{5}{^{\circ}C} interval during the cooling phase. (a) Bragg wavelength shift of Channel $1$ in three cycles at every \SI{5}{^{\circ}C} step. (b) Shift in Bragg wavelengths of Channel $2$, $3$, and $4$ at every \SI{5}{^{\circ}C} step in Cycle $1$.}
\label{fig:Cycle1_temp_Stable}
\end{figure}
\begin{figure}[t!]
\centering\includegraphics[width=12cm]{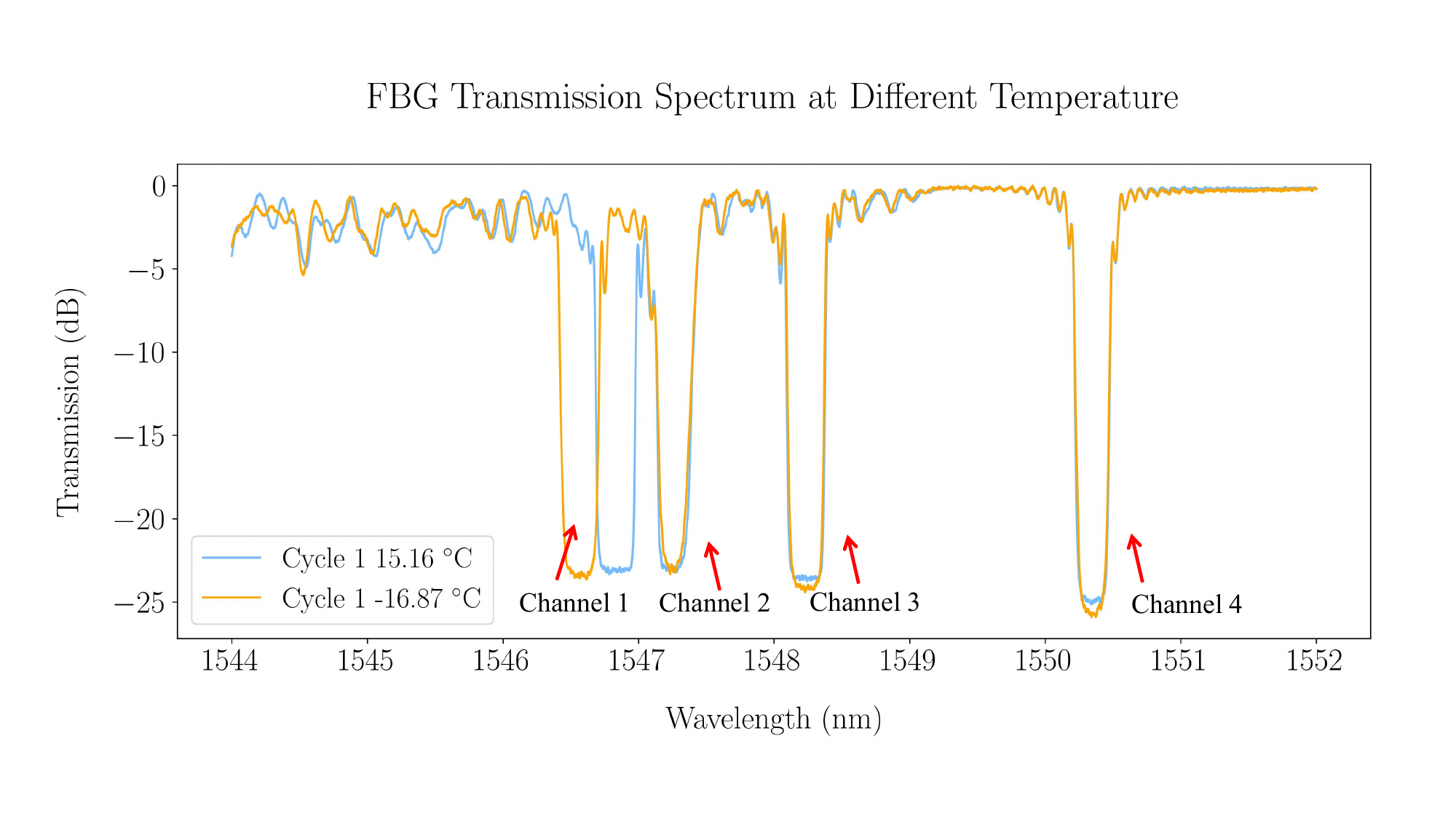}
\caption{Transmission spectrum of the FBG at temperature \SI{15.16}{^{\circ}C} and at the lowest temperature \SI{-16.87}{^{\circ}C} when the temperatures inside the cooling chamber were stabilized in Cycle $1$.}
\label{fig:Channel_full_spectrum}
\end{figure}
\section{Performance study}\label{results_discussions}

In this section, we present the experimental results and conduct a detailed analysis of the behavior of the athermal package. Based on the results, we comment on the bonding process and the behavior of the glue inside the athermal package due to the effects of uncontrolled humidity inside the cooling chamber. 

\subsection{Stable conditions}

\subsubsection{Cycle $1$}

During Cycle $1$, we tracked the center wavelength of each channel as the temperature decreased from room temperature to \SI{-16.87}{^{\circ}C} at every \SI{5}{^{\circ}C}. After each decrement, we allowed the system to stabilize before recording both the transmission spectrum and temperature data. The center wavelength values of each channel at \SI{-16.87}{^{\circ}C} in Cycle $1$ were used as reference points for subsequent wavelength shift comparisons across all three cycles. The measured Bragg wavelength for Channel $1$, $2$, $3$ and $4$ at \SI{-16.87}{^{\circ}C} in Cycle $1$ were \SI{1546.57}{nm}, \SI{1547.26}{nm}, \SI{1548.24}{nm}, \SI{1550.35}{nm}, respectively. 
Figure~\ref{fig:Cycle1_temp_Stable} (a) depicts the wavelength shift of Channel $1$ in three cooling cycles, while Fig.~\ref{fig:Cycle1_temp_Stable} (b) illustrates the Bragg wavelength variations of the compensated channels $2$, $3$, and $4$ (i.e., channels inside the athermal package), at stabilized temperatures recorded at every \SI{5}{^{\circ}C} interval during the cooling phase of Cycle 1. 

In Fig.~\ref{fig:Cycle1_temp_Stable} (a), the uncompensated FBG at Channel $1$, exhibited a total wavelength change of \SI{295}{pm} between room temperature to \SI{-16.87}{^{\circ}C}. Channel $1$ displayed a linear relationship between center wavelength and temperature, with a temperature sensitivity of \SI{8} {pm/^{\circ}C}, obtained by least-square fitting ($R^2 = 0.99939$). The measured temperature sensitivity in the Channel $1$ is almost uniform in different measurement cycles with a low uncertainty of \SI{0.09}{pm}.

\begin{figure}[t]
\centering
\begin{subfigure}{.49\textwidth}
    \centering
    \includegraphics[width=1\linewidth]{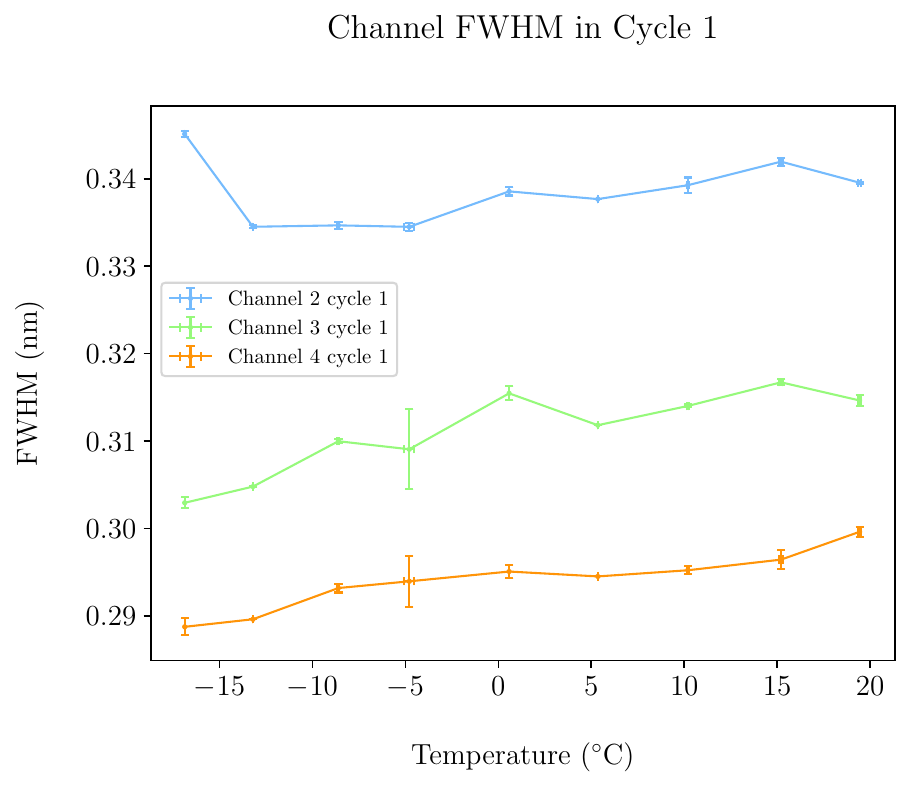}
    \captionsetup{justification=centering, singlelinecheck=false, position=bottom}
    \caption*{\makebox[1.15\linewidth][c]{(a)}}
\end{subfigure}%
\begin{subfigure}{.5\textwidth}
    \centering
    \includegraphics[width=1\linewidth]{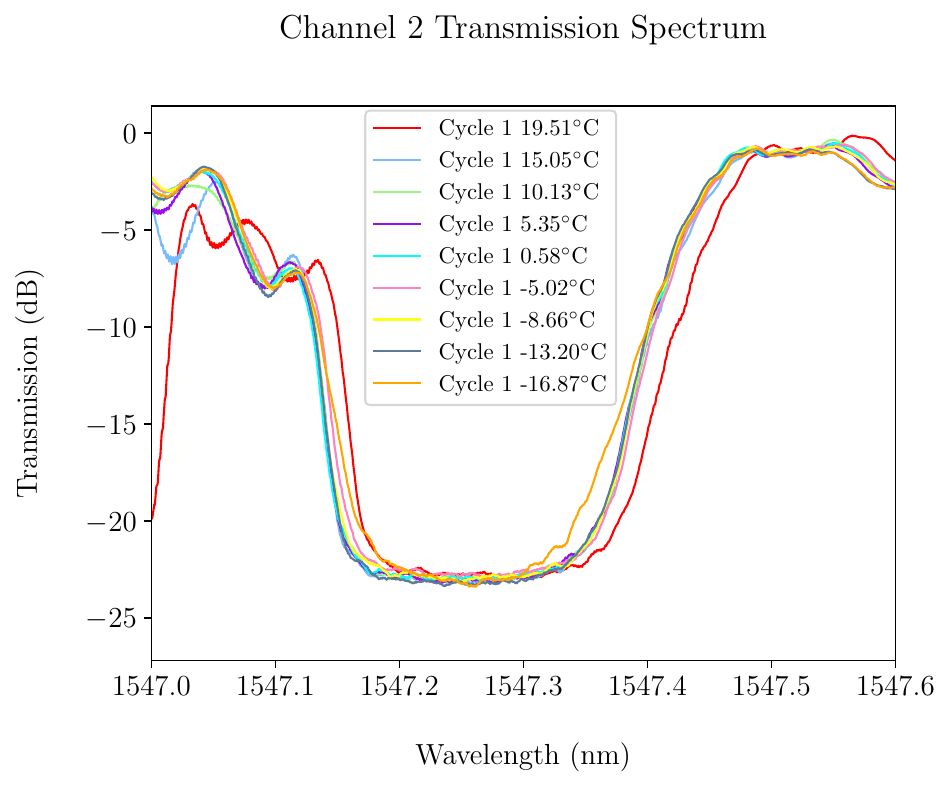}
    \captionsetup{justification=centering, singlelinecheck=false, position=bottom}
    \caption*{\makebox[1.15\linewidth][c]{(b)}}
\end{subfigure}

\vspace{0.5em} 

\begin{subfigure}{.5\textwidth}
    \centering
    \includegraphics[width=1\linewidth]{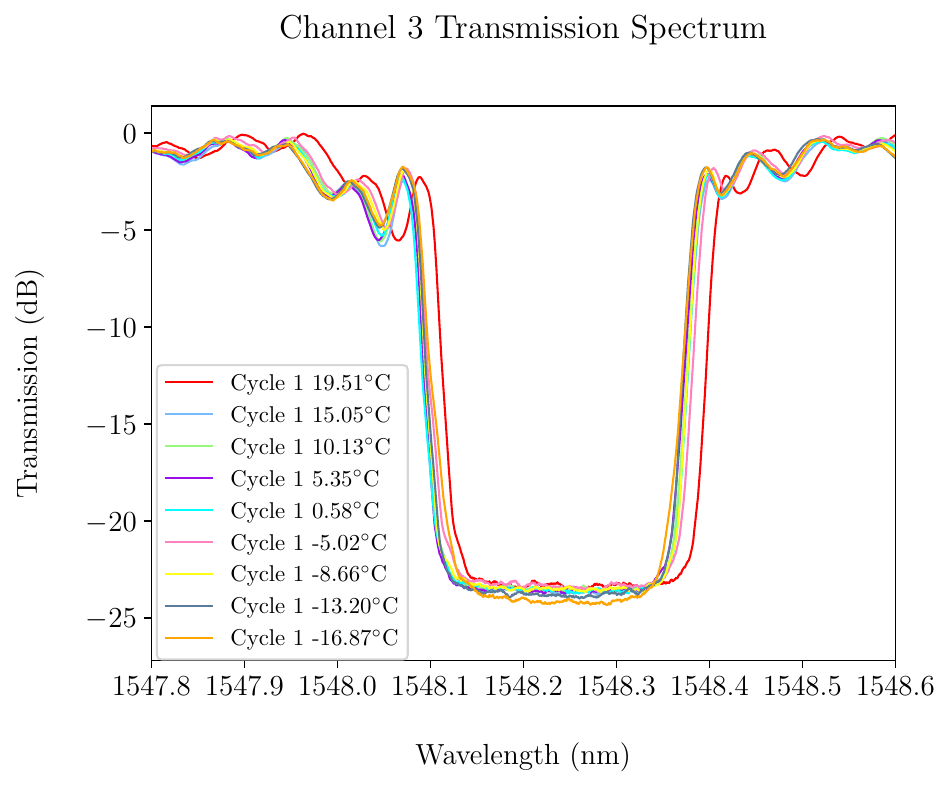}
    \captionsetup{justification=centering, singlelinecheck=false, position=bottom}
    \caption*{\makebox[1.15\linewidth][c]{(c)}}
\end{subfigure}%
\begin{subfigure}{.5\textwidth}
    \centering
    \includegraphics[width=1\linewidth]{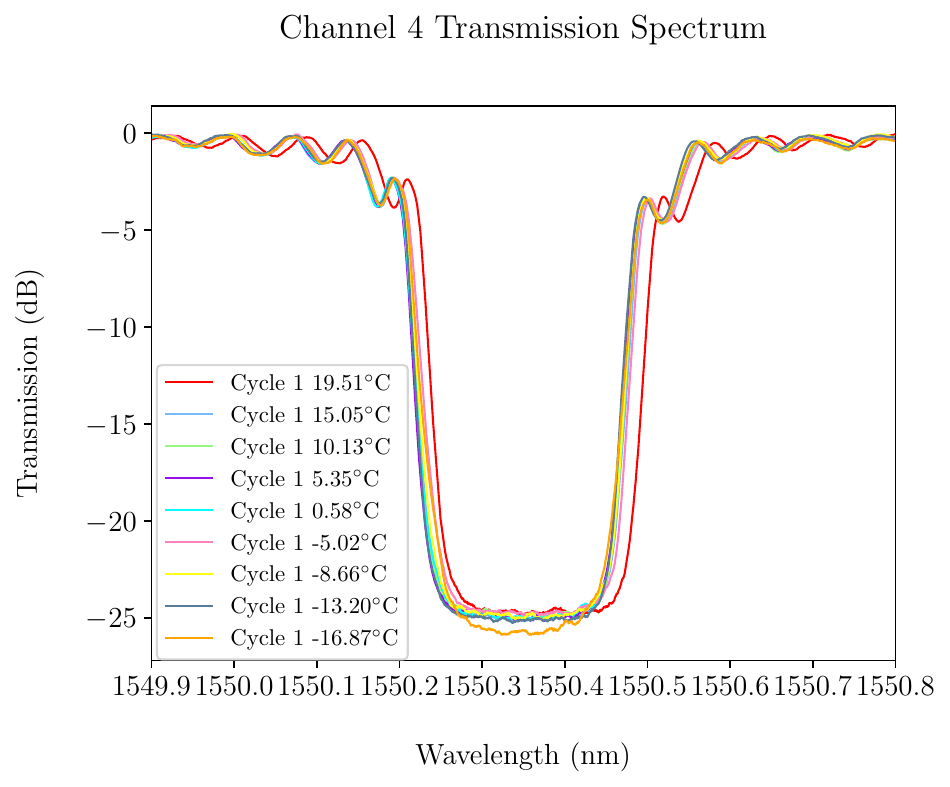}
    \captionsetup{justification=centering, singlelinecheck=false, position=bottom}
    \caption*{\makebox[1.15\linewidth][c]{(d)}}
\end{subfigure}

\caption{Temperature and spectrum shape characterization obtained in the first cooling cycle of the four channels in the temperature stabilized condition. (a) FWHM of the four channels of the FBG when the temperatures inside the cooling chamber were stabilized in Cycle $1$. (b) presents the spectrum shape of Channel $2$. (c) and (d) illustrate the spectrum shape of Channel $3$ and $4$, respectively.}
\label{fig:spectrum_shape}
\end{figure}

In Fig.~\ref{fig:Cycle1_temp_Stable} (b), in Cycle $1$, we observed that Channel $3$ and $4$ exhibited uniform performance, each showing a total wavelength variation of $22$ pm as the temperature changed from \SI{19.44}{^{\circ}C} to \SI{-16.87}{^{\circ}C}. In contrast, Channel $2$ displayed a large wavelength variation of $70$ pm over the same temperature range. However, if the temperature interval from room temperature \SI{19.44}{^{\circ}C} to \SI{15.20}{^{\circ}C} is excluded, Channel $2$'s wavelength variation is comparable to Channel $3$ and $4$, with a wavelength shift of $9$ pm, $10$ pm, and $12$ pm considering all three channels, respectively.  Over the temperature range of \SI{15.16}{^{\circ}C} down to \SI{-16.87}{^{\circ}C}, the athermal package demonstrated good performance, with minimal Bragg wavelength shifts in the channels inside the package and only slight variations in spectral shape, as illustrated in Fig.~\ref{fig:Cycle1_temp_Stable} (b) and Fig.~\ref{fig:Channel_full_spectrum}.
 
 When compared with the uncompensated FBG (Channel $1$), the athermalization factor $F$ of Channel $2$, $3$ and $4$ were $\nicefrac{1}{4}$, $\nicefrac{1}{14}$, $\nicefrac{1}{14}$, respectively, in Cycle $1$, for the temperature range from \SI{19.44}{^{\circ}C} to \SI{-16.87}{^{\circ}C}. The athermalization factor $F$ of Channel $2$, $3$ and $4$ improved to $\nicefrac{1}{30}$, $\nicefrac{1}{26}$, $\nicefrac{1}{22}$, considering the temperature range from \SI{15.20}{^{\circ}C} to \SI{-16.87}{^{\circ}C}.
 
 Figure~\ref{fig:spectrum_shape} shows the full width at half maximum (FWHM) and the spectral shape of Channels $2$, $3$, and $4$ at each stabilized temperature.
 The variation in the FWHM of Channels $2$, $3$ and $4$, as shown in Fig.~\ref{fig:spectrum_shape} (a) is within $\SI{20}{pm}$, the resolution of the OSA. As observed in Fig.~\ref{fig:spectrum_shape} (b), (c) and (d), there are only slight variations in the overall spectral shape of the channels inside the athermal package between $\SI{19.51}{^{\circ}C}$ and $\SI{-16.87}{^{\circ}C}$. However, it is interesting to note that, compared to the spectral shape of Channels $3$ and $4$, Channel $2$ exhibits a slightly asymmetric spectral profile, which is almost consistent over the entire temperature range. The large FWHM of Channel $2$ as compared to Channel $3$ and Channel $4$ is due to a small chirp in the grating, which exhibits as an asymmetry in the spectrum of Channel $2$. We will discuss the asymmetry in Channel $2$ in Section \ref{Analysis}.

\subsubsection {Cycle $2$}
We observed increased wavelength shifts of the FBGs inside the athermal package in Cycle $2$ as compared to that of in Cycle $1$. In Cycle $2$, the Bragg wavelength of Channel $2$ shifted by approximately \SI{79}{pm}, and Channel $3$ and $4$ by \SI{45}{pm} and \SI{43}{pm}, respectively, in the temperature range from \SI{20.09}{^{\circ}C} to \SI{-16.55}{^{\circ}C}. When focusing only on the narrower temperature range from \SI{15.13}{^{\circ}C} to \SI{-16.55}{^{\circ}C} in Cycle $2$, the wavelength shifts in Channel $2$, $3$, and $4$ were \SI{37}{pm}, \SI{39}{pm}, and \SI{35}{pm}, respectively, indicating almost consistent wavelength shifts across the channels in Cycle $2$. However, a degradation in the athermalization factor was observed, decreasing from $\nicefrac{1}{22}$ in Cycle $1$ to $\nicefrac{1}{7}$ in Cycle $2$. We analyze this behavior of the athermal package in Section \ref{Analysis}.

Table~\ref{tab:athermal_perform_23} summarizes the performance of all channels over various temperature ranges in Cycle $1$ and $2$. Cycle $1$ demonstrated the best temperature compensation performance, with a maximum $12$ pm wavelength shift over the range from \SI{15.20}{^{\circ}C} to \SI{-16.87}{^{\circ}C}, equivalent to athermalization factor $F = \nicefrac{1}{22}$.

\begin{table}[ht]
\centering
\caption{The performance of the athermal package in Cycle 1 and 2.}
\label{tab:athermal_perform_23}
\resizebox{\textwidth}{!}{%
  \begin{tabular}{c|c|c|c|c|c|c|c|c}
    \hline
    \multirow{3}{*}{Channel} & \multicolumn{4}{c|}{Cycle 1} & \multicolumn{4}{c}{Cycle 2} \\ \cline{2-9}
    & \multicolumn{2}{c|}{\SI{19.44}{^{\circ}C} to \SI{-16.87}{^{\circ}C}} & \multicolumn{2}{c|}{\SI{15.20}{^{\circ}C} to \SI{-16.87}{^{\circ}C}} &  \multicolumn{2}{c|}{\SI{20.09}{^{\circ}C} to \SI{-16.55}{^{\circ}C}} & \multicolumn{2}{c}{\SI{15.13}{^{\circ}C} to \SI{-16.55}{^{\circ}C}} \\ \cline{2-9}
    & $\Delta\lambda_B$ (pm) & $F$ & $\Delta\lambda_B$ (pm) & $F$ & $\Delta\lambda_B$ (pm) & $F$ & $\Delta\lambda_B$ (pm) & $F$ \\ \hline
    2 & 69 & 1/4 & 9 & 1/30 & 79 & 1/4 & 37 & 1/7 \\ \hline
    3 & 22 & 1/14 & 10 & 1/26 & 45 & 1/7 & 39 & 1/7 \\ \hline
    4 & 22 & 1/14 & 12 & 1/22 & 43 & 1/7 & 35 & 1/8 \\ \hline
  \end{tabular}
}
\end{table}

\subsubsection {Cycle 3}
All channels in Cycle 3 exhibited similar performance to Cycle 2, even though we did not stabilize the cooling chamber at $15^{\circ}$C as we had done in Cycle 2. Moreover, the performance of the athermal package remained stable in Cycle 3, showing no further deterioration.

\begin{figure}[t]
\centering
\begin{subfigure}{.49\textwidth}
    \centering
    \includegraphics[width=1\linewidth]{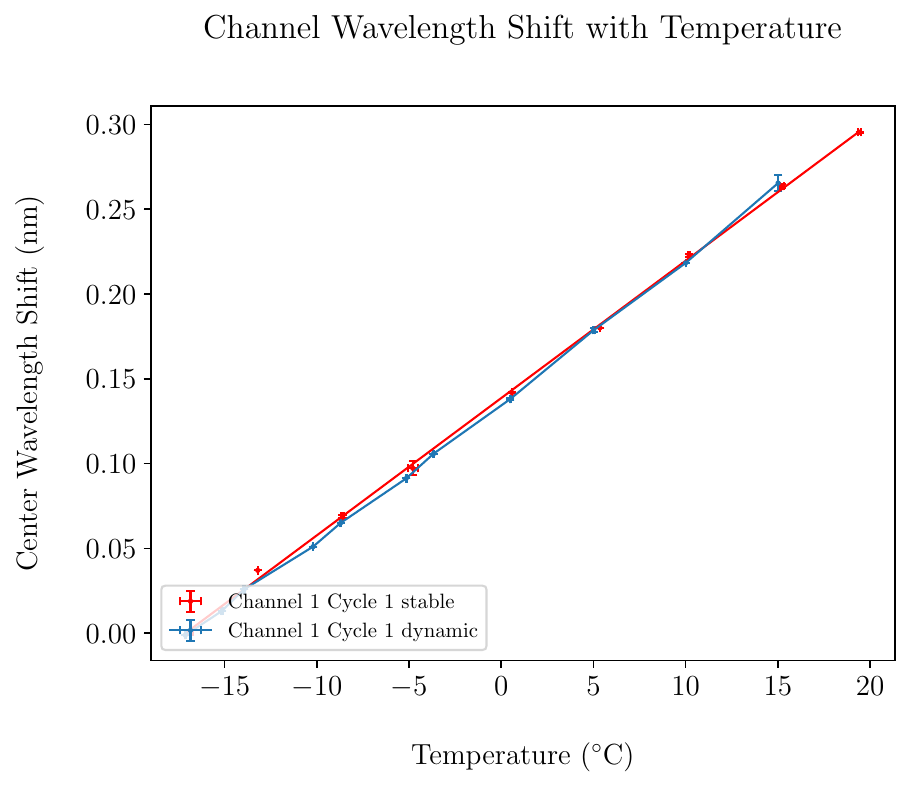}
    \captionsetup{justification=centering, singlelinecheck=false, position=bottom}
    \caption*{\makebox[1.15\linewidth][c]{(a)}}
\end{subfigure}%
\begin{subfigure}{.5\textwidth}
    \centering
    \includegraphics[width=1\linewidth]{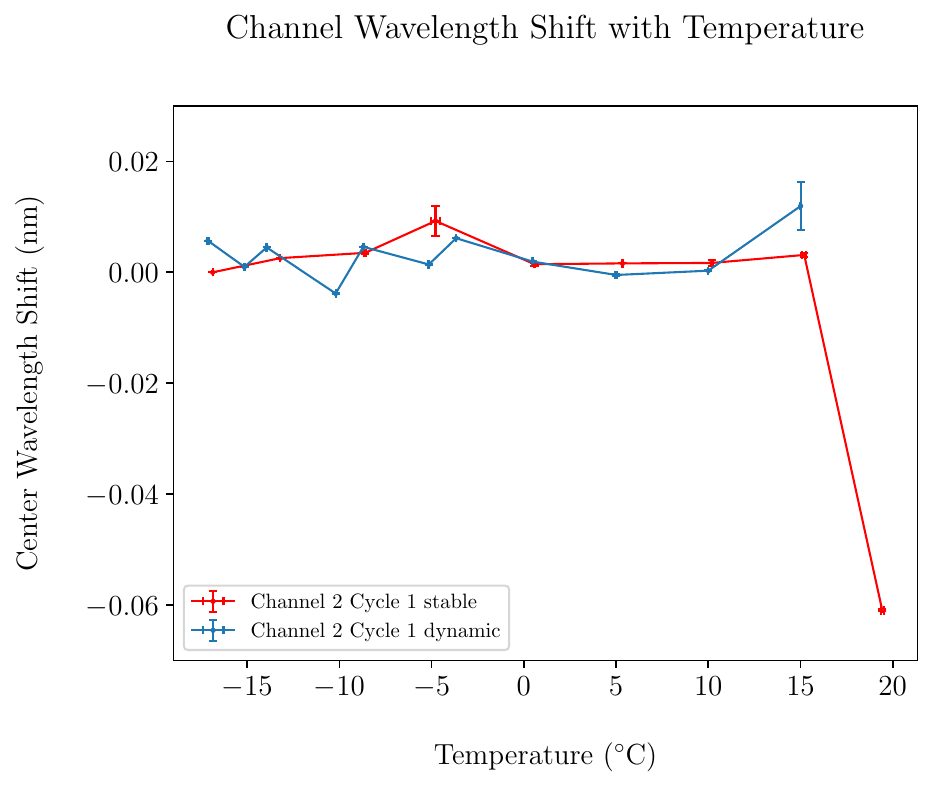}
    \captionsetup{justification=centering, singlelinecheck=false, position=bottom}
    \caption*{\makebox[1.15\linewidth][c]{(b)}}
\end{subfigure}

\vspace{0.5em} 

\begin{subfigure}{.5\textwidth}
    \centering
    \includegraphics[width=1\linewidth]{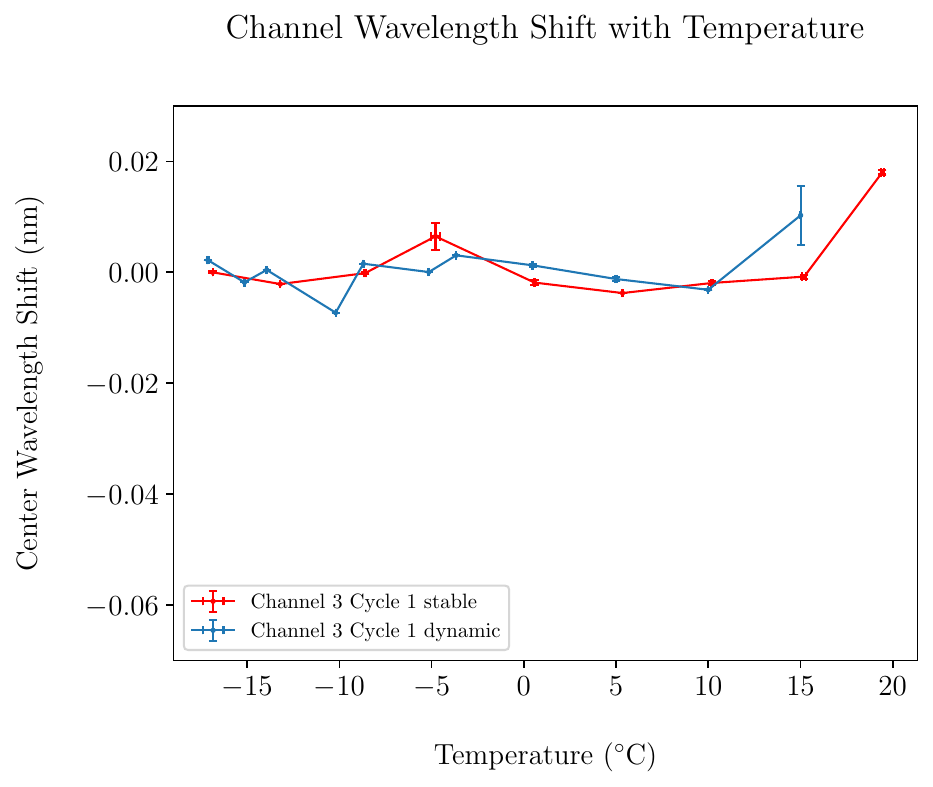}
    \captionsetup{justification=centering, singlelinecheck=false, position=bottom}
    \caption*{\makebox[1.15\linewidth][c]{(c)}}
\end{subfigure}%
\begin{subfigure}{.5\textwidth}
    \centering
    \includegraphics[width=1\linewidth]{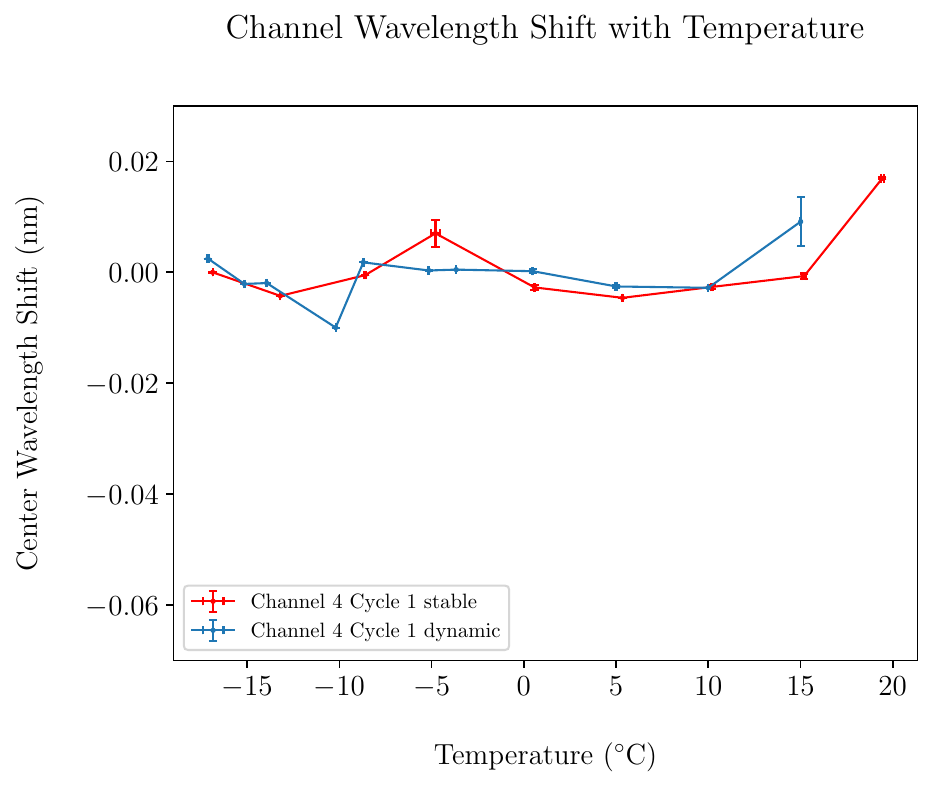}
    \captionsetup{justification=centering, singlelinecheck=false, position=bottom}
    \caption*{\makebox[1.15\linewidth][c]{(d)}}
\end{subfigure}


\caption{Temperature and wavelength shift obtained in the first cooling cycle of the four channels before the temperatures were stabilized, namely the dynamic process. The wavelength shift obtained after the temperatures were stabilized was also displayed. (a) displays the performance of Channel $1$ (b) presents the wavelength shifts of Channel $2$ (c) and (d) illustrate Channel $3$ and $4$, respectively.}
\label{fig:dynamic}
\end{figure}

\subsection{Dynamic conditions}
\subsubsection{Cycle $1$}

In the cooling process of Cycle $1$, the wavelength shifts of all four channels were recorded two times at every \SI{5}{^{\circ}C} interval,  1) when the chamber first reached the intended temperature value (we call it dynamic data), and 2) when the temperature was fully stabilized (stable data). 
In Fig.~\ref{fig:dynamic}, we compare the dynamic data and the stable data to show the dynamic response of the athermal package in Cycle $1$. The Bragg wavelength shift was calculated relative to the wavelength measured at $-16.87$$^{\circ}$C in Cycle $1$. We found that the variation in wavelengths of Channel $2$, $3$, and $4$ remained well below \SI{10}{pm} (for temperature <\SI{15}{^{\circ}C}) as compared to the stabilized temperature condition, whereas, there is a significant mismatch between the dynamic and stable data of the channels inside the athermal package at \SI{15}{^{\circ}C}. This result further supports the observation that the athermal package performs more effectively at temperatures below \SI{15}{^{\circ}C}. Although the rate of change of temperature could not be controlled in the cooling chamber, the athermal package shows a good dynamic response for temperature <\SI{15}{^{\circ}C}.

\subsubsection{Cycle $2$ and Cycle $3$} \label{wavelength_jump}
\begin{figure}[htbp!]
\centering

\begin{subfigure}{.5\textwidth}
    \centering
    \includegraphics[width=1\linewidth]{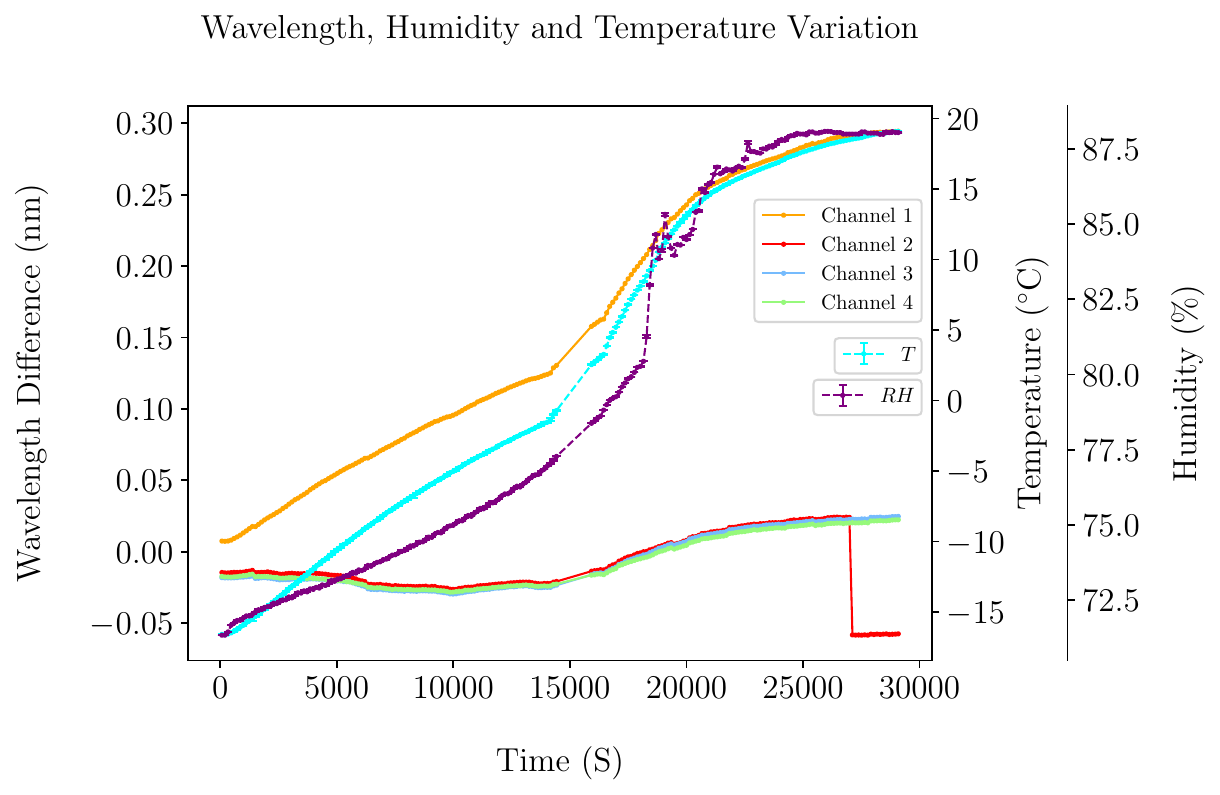}
    \captionsetup{justification=centering, singlelinecheck=false, position=bottom}
    \caption{Cycle 2's heating process}
\end{subfigure}%
\begin{subfigure}{.5\textwidth}
    \centering
    \includegraphics[width=1\linewidth]{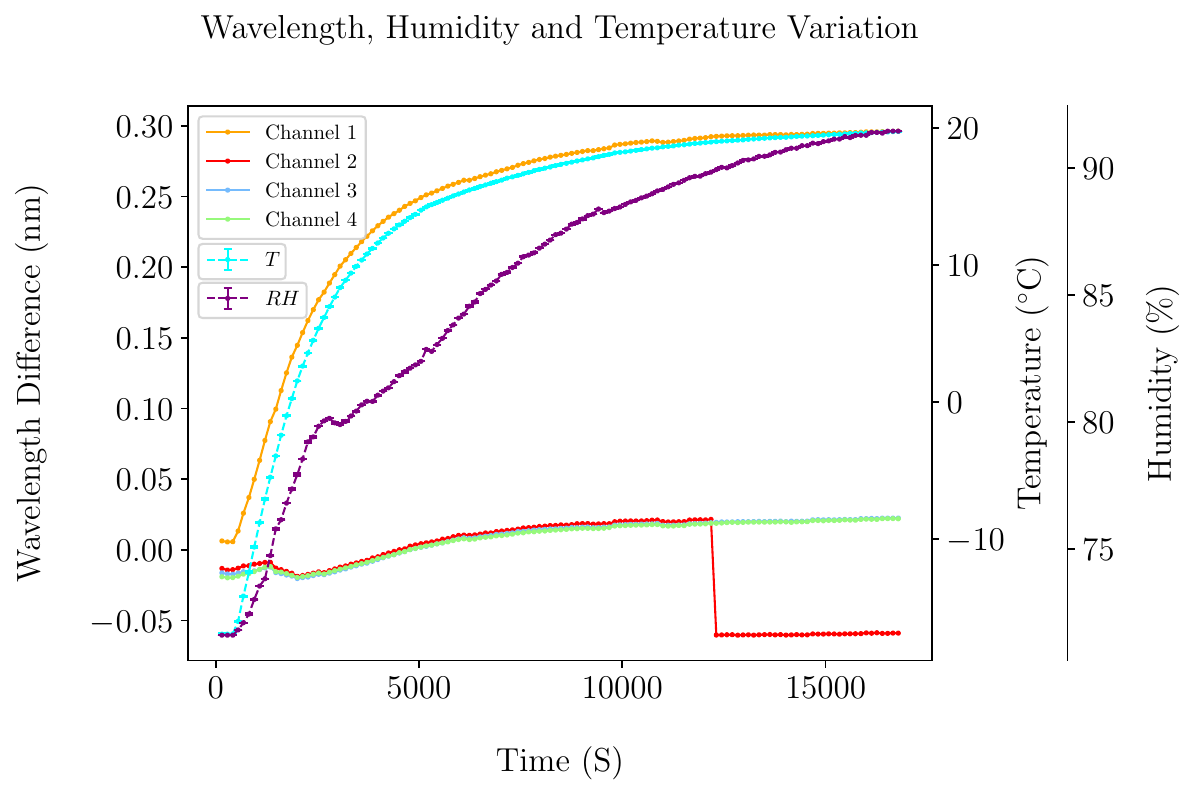}
    \captionsetup{justification=centering, singlelinecheck=false, position=bottom}
    \caption{Cycle 3's heating process}
\end{subfigure}

\vspace{0.5em} 

\begin{subfigure}{.5\textwidth}
    \centering
    \includegraphics[width=1\linewidth]{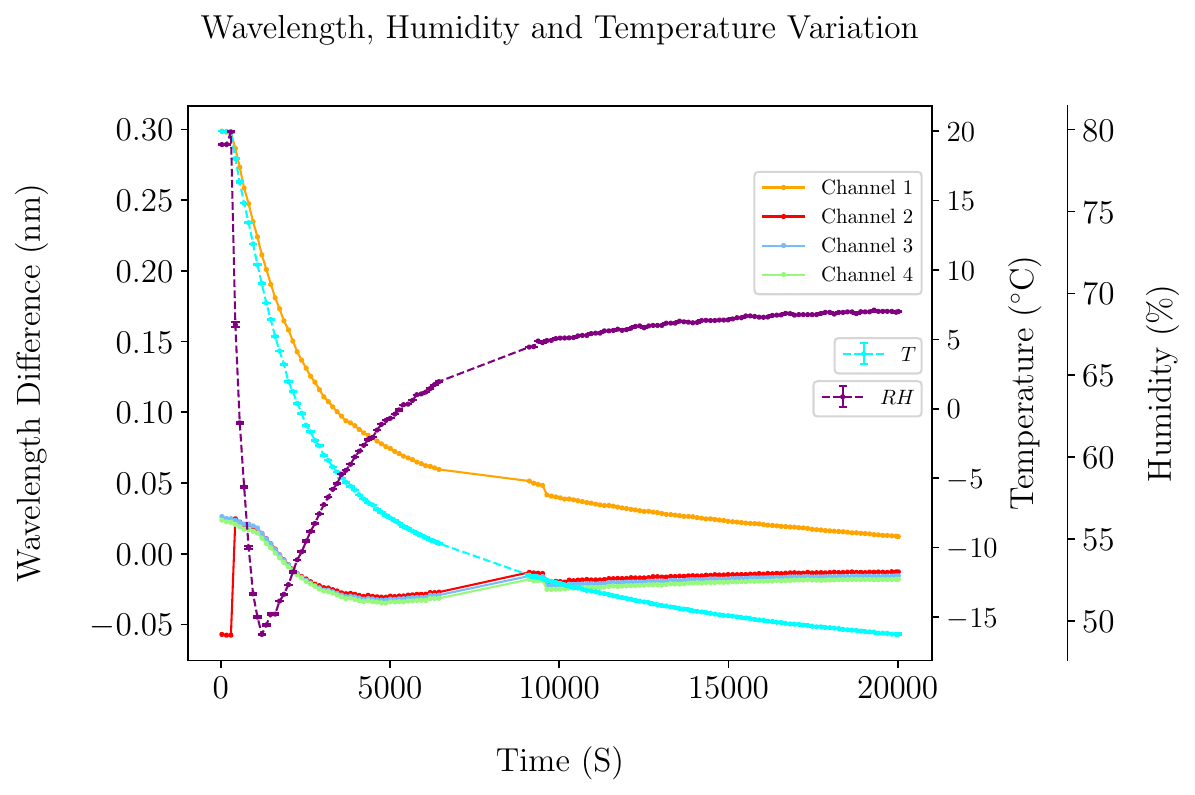}
    \captionsetup{justification=centering, singlelinecheck=false, position=bottom}
    \caption{Cycle 3's cooling process}
\end{subfigure}%

\caption{Shift of Bragg wavelengths of four channels, temperature, and humidity, plotted over time during (a) Cycle $2$'s heating process (b) Cycle 3's heating process and (c) Cycle 3's cooling process.}
\label{fig:Heating Cycle2_3}
\end{figure}

\begin{figure}[htbp!]
\centering

\begin{subfigure}{.5\textwidth}
    \centering
    \includegraphics[width=1\linewidth]{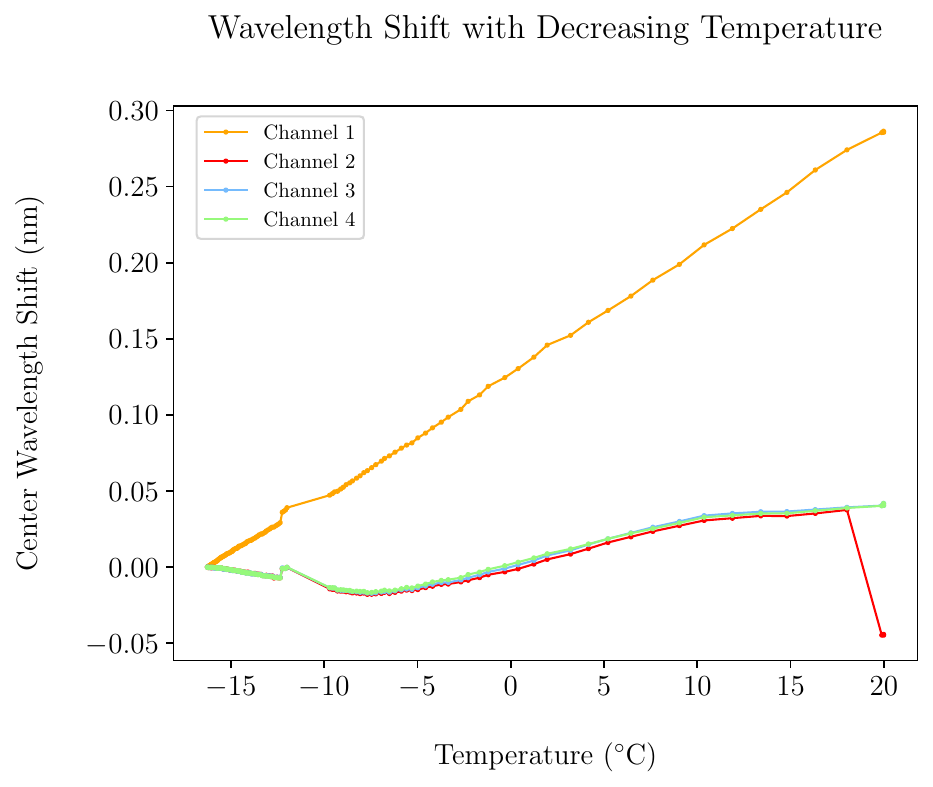}
    \captionsetup{justification=raggedright, singlelinecheck=false, position=top}
    \caption*{\makebox[1.15\linewidth][c]{(a)}}
\end{subfigure}%
\begin{subfigure}{.5\textwidth}
    \centering
    \includegraphics[width=1\linewidth]{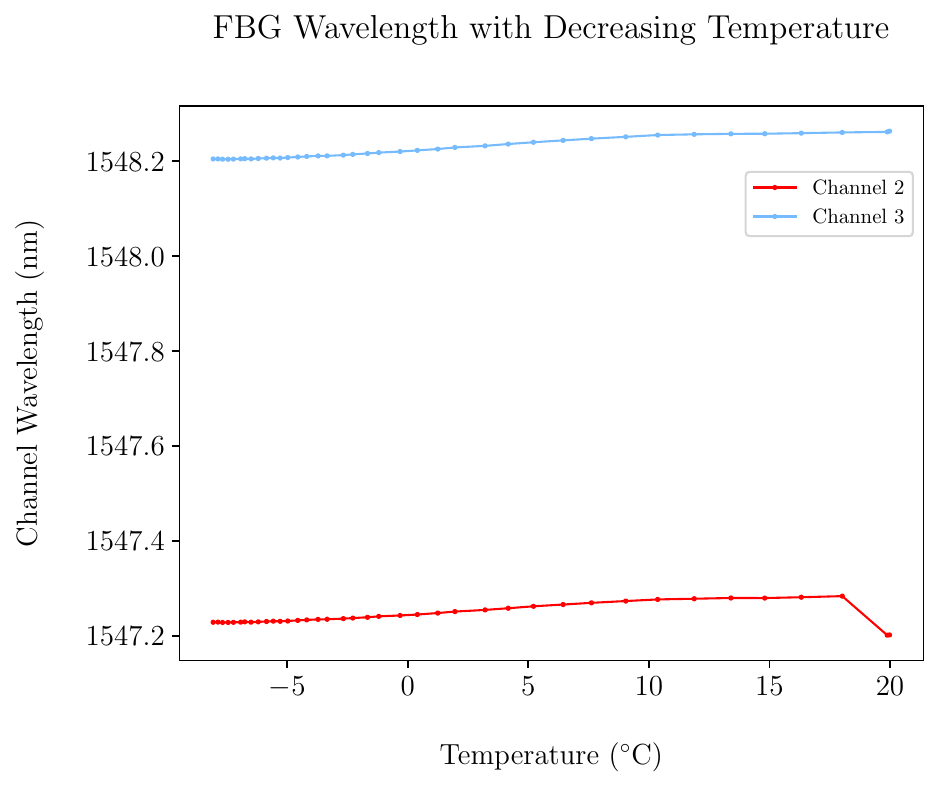}
    \captionsetup{justification=raggedright, singlelinecheck=false, position=top}
    \caption*{\makebox[1.15\linewidth][c]{(b)}}
\end{subfigure}
\caption{Continuous cooling process of Cycle $3$. (a) wavelength shift of four channels (b) Bragg wavelength of Channel $2$ and Channel $3$ in the temperature range of \SI{19.88}{^{\circ}C} to \SI{-5}{^{\circ}C}.}
\label{fig:Cooling_Cycle3_wav_tem_shift}
\end{figure}

In Cycle $2$ and $3$, we analyzed the natural heating process and measured the  
shifts in center wavelength in all channels with respect to the temperature and relative humidity (RH) values, as shown in Fig.~\ref{fig:Heating Cycle2_3}. In these two heating processes, we observed Channel 2 exhibited consistent wavelength jumps (discontinuous wavelength variation) of $82$ pm. During the heating phase of Cycle $2$ and $3$, the wavelength jump of Channel 2 consistently appeared between \SI{18}{^{\circ}C} and \SI{19}{^{\circ}C} (at \SI{18.49}{^{\circ}C} in Cycle $2$ and at \SI{18.95}{^{\circ}C} in Cycle $3$). 
When we compare this observation with the cooling cycle in Cycle 3, we observe, a similar wavelength jump of $82$ pm in Channel $2$ at \SI{18.02}{^{\circ}C}. In Fig.~\ref{fig:Heating Cycle2_3}, we note that the wavelength jump of Channel $2$ is observed when RH changed from $85\%$ to $90\%$ in the heating process and from $80\%$ to $50\%$ in the cooling process. 
 Figure~\ref{fig:Cooling_Cycle3_wav_tem_shift} further illustrates these observations: Fig.~\ref{fig:Cooling_Cycle3_wav_tem_shift} (a) shows the wavelength variation of all four channels in the cooling process of Cycle $3$ while Fig.~\ref{fig:Cooling_Cycle3_wav_tem_shift} (b) depicts the Bragg wavelength of Channel $2$ and $3$ at temperature range \SI{19.88}{^{\circ}C} to \SI{-5}{^{\circ}C}. We realized that Channel $3$ and $4$ performed uniformly and the strain applied to them was less than the temperature decrease effect which need to be compensated. In contrast, Channel $2$ showed the wavelength shift to longer wavelength at \SI{19.89}{^{\circ}C}, which represents that the temperature effect in Channel $2$ was over-compensated at the beginning of the cooling process and later became under-compensated as Channel $3$ and $4$ experienced. We further analyze the implications of these wavelength jumps and the intrinsic character of the athermal package in Section \ref{Analysis}.

\subsection{Hysteresis}
We also analyzed hysteresis effects during the cooling and heating phase of Cycle $3$.
\begin{figure}[htbp!]
\centering
\begin{subfigure}{.49\textwidth}
    \centering
    \includegraphics[width=1\linewidth]{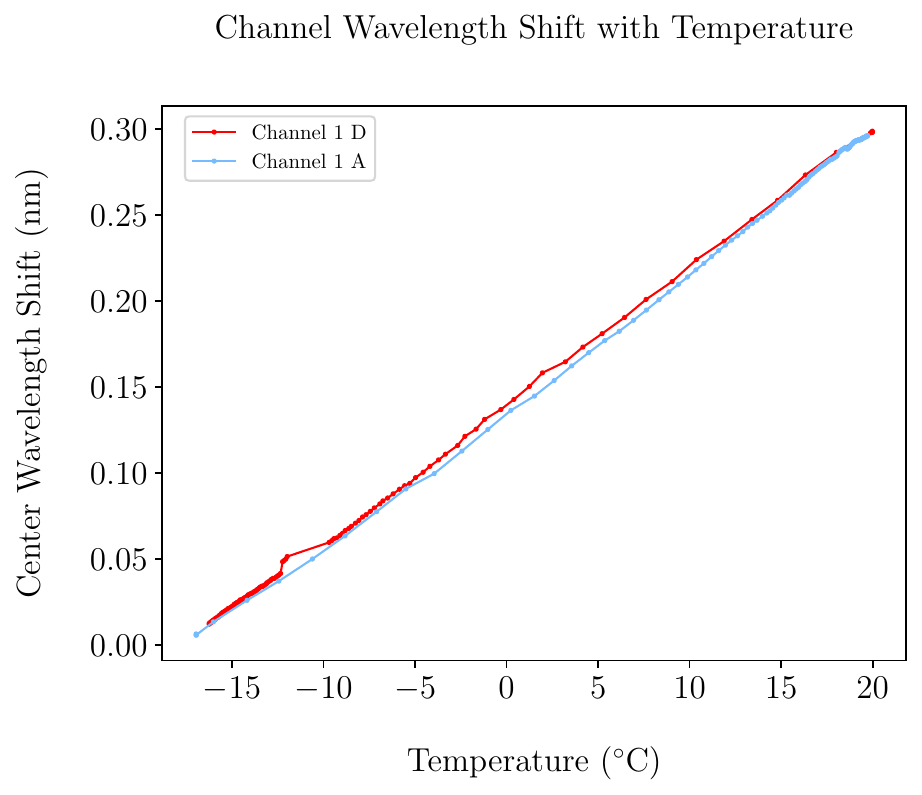}
    \captionsetup{justification=raggedright, singlelinecheck=false, position=top}
    \caption*{\makebox[1.15\linewidth][c]{(a)}}
\end{subfigure}%
\begin{subfigure}{.5\textwidth}
    \centering
    \includegraphics[width=1\linewidth]{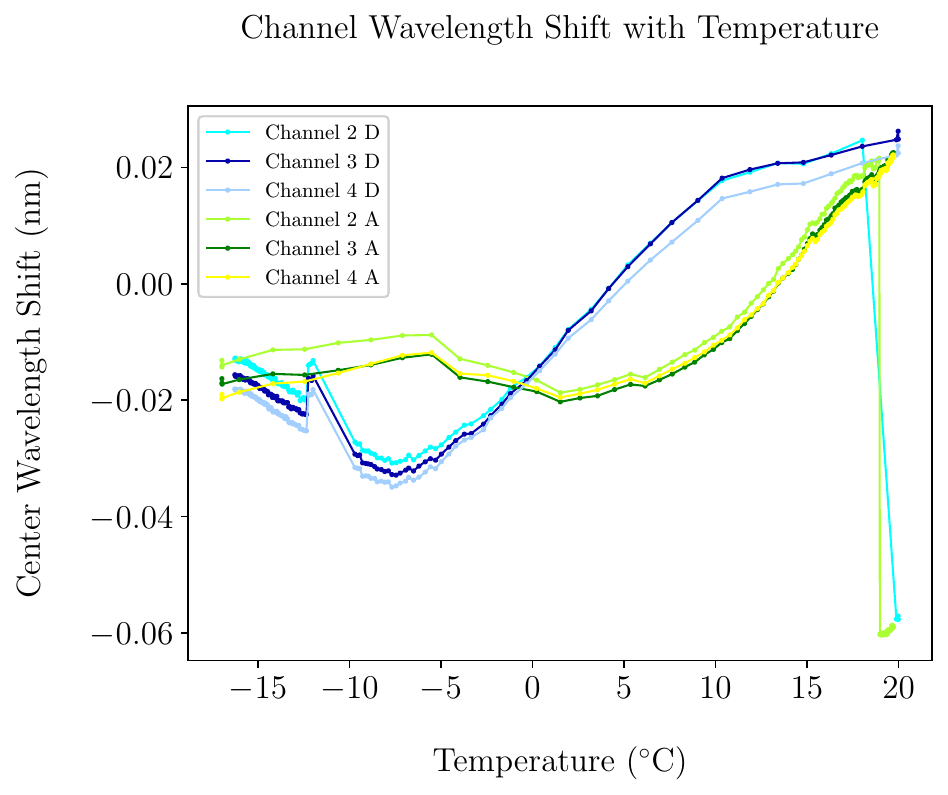}
    \captionsetup{justification=raggedright, singlelinecheck=false, position=top}
    \caption*{\makebox[1.15\linewidth][c]{(b)}}
\end{subfigure}
\caption{Wavelength shift with temperature change obtained in the heating and cooling process of Cycle $3$. The heating process was noted as A and the cooling process of Cycle $3$ was labeled as D. A and D denote the temperature was in ascending or descending order. (a) displays the wavelength shift of Channel $1$. (b) presents the wavelength shift of Channel $2$, $3$, and $4$.}
\label{fig:hystersis}
\end{figure}
Channel $1$ in Fig.~\ref{fig:hystersis} (a) shows no hysteresis while all other channels in Fig.~\ref{fig:hystersis} (b) display hysteresis, demonstrating similar responses across all three channels during the heating and cooling steps within the temperature range of \SI{-17}{^{\circ}C} to \SI{15}{^{\circ}C}. The wavelength jumps over temperature depicted in Fig.~\ref{fig:hystersis} (b) correspond to the wavelength discontinuities described in Section~\ref{wavelength_jump}. Overall, the largest hysteresis effect across the temperature descending and ascending procedure was approximately \SI{24}{pm}. The heating process followed natural warming once the cooling was stopped. Due to uncontrolled humidity, condensed water from the room was manually removed during the heating process. However, any water or dew formed on the athermal package itself was left undisturbed to maintain the integrity of the setup. This likely contributed to the observed lag between the cooling and subsequent heating processes between $\SI{0}{^{\circ}C}$ and $\SI{15}{^{\circ}C}$.  
The discontinuous wavelength variation observed between $\SI{-5}{^{\circ}C}$ and $\SI{-15}{^{\circ}C}$ can be further attributed to a slow response of the temperature sensor (as evident from the discrepancy in temperature behavior between Channel 1, which is located outside the athermal package, and the temperature sensor, as shown in Fig.~\ref{fig:Heating Cycle2_3} (c)) and/or the limited data recording speed of the OSA during this dynamic cooling process. In future studies, the hysteresis effect should be investigated at stabilized temperatures using a controlled heating setup. Given the limitations in the hysteresis study, it is still important to note that all Channels 2, 3, and 4 within the athermal package exhibited a similar pattern of behavior.

\subsection{Performance analysis}\label{Analysis}
The key observations in the performance of the athermal package are, a) among the three channels, Channel 2, which is bonded near the long rod side of the athermal package, initially experiences overcompensation at the start of the cooling process, whereas the other two channels exhibit undercompensation, b) all the channels inside the athermal package undergo larger wavelength deviations between \SI{20}{^{\circ}C} and \SI{15}{^{\circ}C} in Cycle 1, compared to those between \SI{15}{^{\circ}C} and \SI{-17}{^{\circ}C}, c) between \SI{15}{^{\circ}C} and \SI{-17}{^{\circ}C}, all channels behave uniformly while displaying slight undercompensation, and finally, d) the performance of the athermal package deteriorates from Cycle $1$ to Cycle $2$, but no further degradation is observed in Cycle $3$. 

In this section, we analyze our results to understand the key issues and also discuss the potential strategies to mitigate them in future studies.

1) Asymmetric design: 
In the temperature range of $\SI{20}{^{\circ}C}$ to $\SI{15}{^{\circ}C}$, we observed that while Channel $3$ and $4$ showed a wavelength shift to the shorter wavelength exhibiting under-compensation, Channel $2$ showed a wavelength shift to the longer wavelength exhibiting over-compensation. This contrasting behavior of Channel $2$ with Channel $3$ and $4$ can be attributed to the asymmetric design of the athermal package, as one of the compensating ends experiences greater displacement with temperature variations than the other (with decrease in temperature, contraction in D $>$ contraction in C), as demonstrated in the modeling results shown in Fig.~\ref{fig:femresults}. During the beginning of the cooling process, the longer side of the aluminium rod may exhibit larger pulling force as compared to the other side, resulting in an over-compensation for Channel $2$, while Channel $3$ and $4$ remain under-compensated. 
However, it is unclear why this asymmetric behavior is not observed below \SI{15}{^{\circ}C}. In general, this effect effectively limits the usable length of the fiber within the athermal package to less than \SI{110}{mm}. However, as long as the maximum wavelength deviation among all channels remains within acceptable limits in a certain working temperature regime, it is still possible to utilize the entire fiber length. 
In our future studies, we will conduct an in-depth investigation to calibrate the behavior of different sections of the fiber within the athermal package by incorporating a long-length single FBG ($\sim$$\SI{100}{mm}$). A change in the spectral shape (FWHM, spectral asymmetry) of the FBG with temperature can provide more information on the non-uniform deformation along the length of the fiber. It is to be noted that, in a real on-sky test scenario where the OH filter is a single, long-length multinotch filter, any chirp will increase the FHWM and therefore can render the filtered part unusable in astronomical observations.
The complete details of the design wavelength of the filters and their corresponding FWHM are found in \cite{Trinh_2013}. 
The allowed increase in the FWHM of the filter lines depends on the resolution of the spectrograph. For example, \cite{Ellis2020} emphasizes that if the OH lines are filtered at a resolution of ($\Delta\lambda\approx$$\SI{180}{pm}$), they occupy $<$ $\SI{0.5}{pixels}$ on the spectrograph and do not spread over several pixels so that the filtered part remains usable for measurement from astronomical observations.

2) Glue performance: While the asymmetric design of the athermal package may be partly responsible for the contrasting wavelength deviations of Channel $2$ compared to Channel $3$ and $4$ at the beginning of the cooling process, we also observed larger wavelength shifts in all the channels between \SI{20}{^{\circ}C} and \SI{15}{^{\circ}C} in Cycle $1$ and only larger wavelength shift in Channel $2$ in the later cycles, 
We suspect that moisture-related effects on the glue also contributed to these inconclusive results\cite{montero2014influence, de1992effects}. The deterioration of the glue and, therefore, the impact on the bonding of the fiber can affect both the linearity and repeatability of wavelength shift measurements, potentially compromising the accuracy of temperature compensation\cite{Kuang}.
In our experiments, the glue used was not humidity-resistant. It is suspected that the glue's interactions with high levels of humidity at and around the dew point (at that time of the experiment, the dew point was between \SI{20}{^{\circ}C} and \SI{15}{^{\circ}C}) contributed to the unexplained behavior of the athermal package within the specified temperature range. The wavelength jump observed in Section~\ref{wavelength_jump} could have been caused by the moisture absorption of the glue which resulted in a reduced performance.

Furthermore, as for the design of the experiments, Cycle 1 required a longer duration (with stabilization at every \SI{5}{^{\circ}C} for a day) compared to Cycles $2$ and $3$. The prolonged exposure to moisture may have altered the properties of the glue \cite{montero2014influence, de1992effects}, potentially contributing to the performance degradation observed in Cycle $2$. 

The performance of the athermal package can be enhanced by selecting a glue with better tolerance to humidity variations, a critical factor in astronomical applications.

3) Non-linear expansion: We also considered understanding the aforementioned effect in light of the non-linear thermal expansion of aluminum within the athermal package. As the temperature decreases, the CTE of aluminum exhibits a slight reduction \cite{Wilson2002}. If non-linear thermal expansion were a dominant factor, we would anticipate results contrary to our observations, i.e., larger strain at higher temperatures would result in smaller wavelength shifts in the channels as the temperature decreases, leading to better compensation at higher temperatures than lower temperatures. Therefore, we can conclude that the non-linearity in the thermal expansion of the materials was not the primary cause of the observed increase in wavelength shifts of all the channels in Cycle $1$ in the temperature between \SI{20}{^{\circ}C} and \SI{15}{^{\circ}C}. However, we would model the non-linearity in thermal expansion coefficients in our future studies to obtain more insight into this.

Further, we comment on a fourth aspect, that may also have impacted the overall performance of the athermal package.

4) Fiber bonding process: 
After completing the fiber bonding process (Section~\ref{assembly}), 
we observed non-uniform shifts among the channels in the athermal package when comparing their wavelengths before and after the bonding. This is illustrated by Fig.~\ref{fig:FBG_transmission_release_spec}, where the Bragg wavelengths of Channel 
$2$, $3$ and $4$ shifted by 
$7$ pm, $27$ pm, and $22$ pm, respectively, after the FBGs were bonded to the package compared to the uninstalled, room-temperature condition. The asymmetric spectral profile of Channel $2$ as observed in Fig.~\ref{fig:spectrum_shape} (b) suggests a small chirp in the grating.
The exact cause of these discrepancies is unclear; however, handling the fiber during the bonding process likely introduced the error. Channel $2$, located near the longer rod side of the athermal package (Fig.~\ref{fig:athermal_pack_layout}), may have experienced a non-uniform strain during bonding, leading to the observed non-uniform shift and chirp, when compared with the other two channels. Therefore, an improvement to the athermal package design is needed so that post-inscription handling of the fiber is avoided.  

In the second-generation \cite{Patent} athermal package, we have introduced an improved design by incorporating an open window in the cylindrical athermal package. In this new design, the fiber is bonded first, followed by filter inscription. This modification eliminates the need for post-inscription bonding/handling, which can introduce non-uniform strain and thus prevent the use of longer filters for astronomical applications.
We also observed a significant hysteresis between the cooling and subsequent heating cycles in Cycle 3 (Fig.~\ref{fig:hystersis} (b)). A likely cause was the formation of dew on the athermal package during the heating process. This will warrant an upgraded packaging system so that dew formation is prevented. 

We summarize that within the temperature range from \SI{-17}{^{\circ}C} to \SI{15}{^{\circ}C}, temperature compensation remained consistent across all channels, demonstrating the potential of this athermal package for long-length FBGs for astronomy, with provisions for further improvements. For enhanced performance, we will address the following aspects in our future communication (Part 3 in preparation): 1) calibration of the effect of temperature variations along the fiber length within the athermal package to better understand non-uniform deformation, 2) use of humidity-resistant glue, and 3) fabrication of filters in a post-bonded fiber-athermal package unit using the second-generation athermal package.


%


\section{Conclusions}\label{conclusions}
We developed a prototype of a self-compensating athermal package that can house a  \SI{110}{mm} long optical fiber. 
The performance of the package was studied over three cycles, from \SI{-17}{^{\circ}C} to \SI{20}{^{\circ}C}. However, due to the influence of humidity on the glue used to bond the fiber to the athermal package, and also, the non-uniform bonding process of one end of the fiber, the results for the temperature range from \SI{15}{^{\circ}C} to \SI{20}{^{\circ}C} were inconclusive. For the narrower temperature range of \SI{-17}{^{\circ}C} to \SI{15}{^{\circ}C}, we observed a maximum Bragg wavelength shift of 12 pm in the channels inside the athermal package. This corresponds to a temperature athermalization factor $F$ of $\nicefrac{1}{22}$, compared to the uncompensated channel (with a temperature sensitivity of \SI{8}{pm/^{\circ}C}) over a temperature range of \SI{35}{^{\circ}C}. We also demonstrate that the temperature compensation across all three channels, spanning over a fiber length of $\sim$ \SI{110}{mm}, remains highly consistent in the above-mentioned temperature range, demonstrating its suitability for long-length FBG filters, such as multi-channel FBG filters or chirped FBGs for astronomical applications. In the next-generation design, we introduce an FBG inscription window in the athermal package, allowing filters to be inscribed after bonding the fiber to the package, to eliminate unwanted chirp induced during the bonding process and improve wavelength precision. Additionally, we will use a humidity-resistant glue to address the performance limitations observed in the current prototype.



\begin{backmatter}
\bmsection{Funding}
This work is supported by DFG (Deutsche Forschungsgemeinschaft project no.~455425131).

\bmsection{Acknowledgment}
The OH lines observation was based on data from the CAHA Archive at CAB (INTA-CSIC). The CAHA Archive is part of the Spanish Virtual Observatory project funded by MCIN/AEI/10.13039/501100011033 through grant PID2020-112949GB-I00.

\bmsection{Disclosures} The authors declare no conflicts of interest.

\bmsection{Data availability} Data underlying the results presented in this paper are not publicly available at this time but may be obtained from the authors upon reasonable request.

\end{backmatter}

\bibliography{sample}

\begin{thebibliography}{10}
\newcommand{\enquote}[1]{``#1''}

\bibitem{bilodeau1995all}
F.~Bilodeau, D.~Johnson, S.~Theriault, \emph{et~al.}, \enquote{An all-fiber dense wavelength-division multiplexer/demultiplexer using photoimprinted {B}ragg gratings,} {\protect\JournalTitle{IEEE Photonics Technology Letters}} \textbf{7}, 388--390 (1995).

\bibitem{guy1995single}
M.~Guy, J.~Taylor, and R.~Kashyap, \enquote{Single-frequency erbium fibre ring laser with intracavity phase-shifted fibre {B}ragg grating narrowband filter,} {\protect\JournalTitle{Electronics Letters}} \textbf{31}, 1924--1925 (1995).

\bibitem{eggleton1999electrically}
B.~J. Eggleton, J.~A. Rogers, P.~S. Westbrook, and T.~A. Strasser, \enquote{Electrically tunable power efficient dispersion compensating fiber {B}ragg grating,} {\protect\JournalTitle{IEEE Photonics Technology Letters}} \textbf{11}, 854--856 (1999).

\bibitem{Kok2024}
S.~P. Kok, Y.~I. Go, X.~Wang, and M.~L.~D. Wong, \enquote{Advances in fiber {Bragg} grating {(FBG)} sensing: A review of conventional and new approaches and novel sensing materials in harsh and emerging industrial sensing,} {\protect\JournalTitle{IEEE Sensors Journal}} \textbf{24}, 29485--29505 (2024).

\bibitem{Theo2024}
A.~Theodosiou, \enquote{Recent advances in fiber {B}ragg grating sensing,} {\protect\JournalTitle{Sensors}} \textbf{24}, 532--539 (2024).

\bibitem{Meinel}
A.~B. {Meinel}, II, \enquote{{OH Emission Bands in the Spectrum of the Night Sky. II.}} {\protect\JournalTitle{\apj}} \textbf{112}, 120--130 (1950).

\bibitem{Trinh_2013}
C.~Q. Trinh, S.~C. Ellis, J.~Bland-Hawthorn, \emph{et~al.}, \enquote{{GNOSIS}: The first instrument to use fiber {B}ragg gratings for {OH} suppression,} {\protect\JournalTitle{The Astronomical Journal}} \textbf{145}, 51--64 (2013).

\bibitem{Ellis2020}
S.~C. Ellis, J.~Bland-Hawthorn, J.~S. Lawrence, \emph{et~al.}, \enquote{First demonstration of {OH} suppression in a high-efficiency near-infrared spectrograph,} {\protect\JournalTitle{Monthly Notices of the Royal Astronomical Society}} \textbf{492}, 2796--2806 (2020).

\bibitem{Diab:21}
M.~Diab, A.~Tripathi, J.~Davenport, \emph{et~al.}, \enquote{Simulations of mode-selective photonic lanterns for efficient coupling of starlight into the single-mode regime,} {\protect\JournalTitle{Appl. Opt.}} \textbf{60}, D9--D14 (2021).

\bibitem{Davenport:21}
J.~J. Davenport, M.~Diab, P.~J. Deka, \emph{et~al.}, \enquote{Photonic lanterns: a practical guide to filament tapering,} {\protect\JournalTitle{Opt. Mater. Express}} \textbf{11}, 2639--2649 (2021).

\bibitem{Martin}
M.~M. Roth, Z.~Zhang, K.~V. Madhav, and J.~Fiebrandt, \enquote{{FBG development for OH suppression at innoFSPEC Potsdam (Conference Presentation)},} in \emph{Advances in Optical and Mechanical Technologies for Telescopes and Instrumentation III,}  vol. 10706 R.~Navarro and R.~Geyl, eds., International Society for Optics and Photonics (SPIE, 2018), p. 1070605.

\bibitem{Diab_mnras}
M.~Diab, A.~N. Dinkelaker, J.~Davenport, \emph{et~al.}, \enquote{Starlight coupling through atmospheric turbulence into few-mode fibres and photonic lanterns in the presence of partial adaptive optics correction,} {\protect\JournalTitle{Monthly Notices of the Royal Astronomical Society}} \textbf{501}, 1557--1567 (2020).

\bibitem{Davenport:21_2}
J.~J. Davenport, M.~Diab, K.~Madhav, and M.~M. Roth, \enquote{Optimal {SMF} packing in photonic lanterns: comparing theoretical topology to practical packing arrangements,} {\protect\JournalTitle{J. Opt. Soc. Am. B}} \textbf{38}, A7--A14 (2021).

\bibitem{Julian}
J.~Rypalla, S.~Vješnica, K.~Madhav, \emph{et~al.}, \enquote{{On the large quantity fabrication and reproducibility of all-fiber photonic lanterns},} in \emph{Advances in Optical and Mechanical Technologies for Telescopes and Instrumentation VI,}  vol. 13100 R.~Navarro and R.~Jedamzik, eds., International Society for Optics and Photonics (SPIE, 2024), p. 131006P.

\bibitem{ellis2008case}
S.~Ellis and J.~Bland-Hawthorn, \enquote{The case for {OH} suppression at near-infrared wavelengths,} {\protect\JournalTitle{Monthly Notices of the Royal Astronomical Society}} \textbf{386}, 47--64 (2008).

\bibitem{Maihara_1993}
T.~Maihara, F.~Iwamuro, T.~Yamashita, \emph{et~al.}, \enquote{Observations of the {OH} airglow emission,} {\protect\JournalTitle{Publications of the Astronomical Society of the Pacific}} \textbf{105}, 940--944 (1993).

\bibitem{dauphin2024hydroxyl}
F.~Dauphin, A.~Petric, {\'E}.~Artigau, \emph{et~al.}, \enquote{Hydroxyl lines and moonlight: a high spectral resolution investigation of {NIR} skylines from {M}aunakea to guide {NIR} spectroscopic surveys,} {\protect\JournalTitle{arXiv:2412.05473}}  (2024).

\bibitem{SKAAR2001}
J.~Skaar, L.~Wang, and T.~Erdogan, \enquote{On the synthesis of fiber {B}ragg gratings by layer peeling,} {\protect\JournalTitle{IEEE Journal of Quantum Electronics}} \textbf{37}, 165--173 (2001).

\bibitem{Skaar2002}
J.~Skaar and R.~Feced, \enquote{Reconstruction of gratings from noisy reflection data,} {\protect\JournalTitle{J. Opt. Soc. Am. A}} \textbf{19}, 2229--2237 (2002).

\bibitem{Buryak:03}
A.~Buryak, K.~Kolossovski, and D.~Stepanov, \enquote{Optimization of refractive index sampling for multichannel fiber {B}ragg gratings,} {\protect\JournalTitle{IEEE Journal of Quantum Electronics}} \textbf{39}, 91--98 (2003).

\bibitem{Bland-Hawthorn:08}
J.~Bland-Hawthorn, A.~Buryak, and K.~Kolossovski, \enquote{Optimization algorithm for ultrabroadband multichannel aperiodic fiber {B}ragg grating filters,} {\protect\JournalTitle{J. Opt. Soc. Am. A}} \textbf{25}, 153--158 (2008).

\bibitem{Liu}
Y.~Liu, J.-J. Pan, C.~Gu, \emph{et~al.}, \enquote{{Novel fiber Bragg grating fabrication method with high-precision phase control},} {\protect\JournalTitle{Optical Engineering}} \textbf{43}, 1916 -- 1922 (2004).

\bibitem{Petermann2002}
I.~Petermann, B.~Sahlgren, S.~Helmfrid, \emph{et~al.}, \enquote{Fabrication of advanced fiber {B}ragg gratings by use of sequential writing with a continuous-wave ultraviolet laser source,} {\protect\JournalTitle{Appl. Opt.}} \textbf{41}, 1051--1056 (2002).

\bibitem{Gbadebo2018}
A.~A. Gbadebo, E.~G. Turitsyna, and J.~A.~R. Williams, \enquote{Fabrication of precise aperiodic multichannel fibre {B}ragg grating filters for spectral line suppression in hydrogenated standard telecommunications fibre,} {\protect\JournalTitle{Opt. Express}} \textbf{26}, 1315--1323 (2018).

\bibitem{Stepanov}
D.~Y. Stepanov, S.~Surve, and S.~A. Balon, \enquote{Pitch control in fabrication of {B}ragg gratings,} in \emph{Conference on Lasers and Electro-Optics/Quantum Electronics and Laser Science Conference and Photonic Applications Systems Technologies,}  (Optica Publishing Group, 2006), p. CTuY1.

\bibitem{Gagne:08}
M.~Gagn\'{e}, L.~Bojor, R.~Maciejko, and R.~Kashyap, \enquote{Novel custom fiber {B}ragg grating fabrication technique based on push-pull phase shifting interferometry,} {\protect\JournalTitle{Opt. Express}} \textbf{16}, 21550--21557 (2008).

\bibitem{BuryakPatent}
A.~{Buryak}, \enquote{{Method for designing optimised multi-channel grating structures},} U.S.patent 8,194,320B2 (5 June 2012).

\bibitem{Rahman:20}
A.~Rahman, K.~Madhav, and M.~M. Roth, \enquote{Complex phase masks for {OH} suppression filters in astronomy: part {I}: design,} {\protect\JournalTitle{Opt. Express}} \textbf{28}, 27797--27807 (2020).

\bibitem{Rahman:23}
A.~Rahman, T.~Siefke, K.~Madhav, \emph{et~al.}, \enquote{Design and fabrication of a novel phase mask to inscribe fiber {B}ragg gratings for astronomical applications,} in \emph{CLEO 2023,}  (Optica Publishing Group, 2023), p. SF1H.3.

\bibitem{Luo}
X.~Luo, A.~Rahman, K.~Madhav, \emph{et~al.}, \enquote{{Novel phase masks with overlapping regions to fabricate fiber Bragg gratings for filtering sky emission lines},} in \emph{Advances in Optical and Mechanical Technologies for Telescopes and Instrumentation VI,}  vol. 13100 R.~Navarro and R.~Jedamzik, eds., International Society for Optics and Photonics (SPIE, 2024), p. 1310068.

\bibitem{Hernandez}
E.~Hernandez, A.~G{\"u}nther, S.~Vješnica, \emph{et~al.}, \enquote{{Preliminary results of the Potsdam Arrayed Waveguide Spectrograph},} in \emph{Ground-based and Airborne Instrumentation for Astronomy X,}  vol. 13096 J.~J. Bryant, K.~Motohara, and J.~R.~D. Vernet, eds., International Society for Optics and Photonics (SPIE, 2024), p. 130960L.

\bibitem{Liske}
{Liske, Jochen, and E-ELT Project Science Team}, \enquote{{E-ELT Programme: Top Level Requirements for ELT-HIRES},} \url{https://www.eso.org/sci/facilities/eelt/docs/ESO-204697_1_Top_Level_Requirements_for_ELT-HIRES.pdf}.

\bibitem{Kuang}
Y.~{Kuang}, Y.~{Guo}, L.~{Xiong}, and W.~{Liu}, \enquote{{Packaging and temperature compensation of fiber Bragg grating for strain sensing: A survey},} {\protect\JournalTitle{Photonic Sensors}} \textbf{8}, 320--331 (2018).

\bibitem{yoffe1995passive}
G.~Yoffe, P.~A. Krug, F.~Ouellette, and D.~Thorncraft, \enquote{Passive temperature-compensating package for optical fiber gratings,} {\protect\JournalTitle{Applied optics}} \textbf{34}, 6859--6861 (1995).

\bibitem{iwashima1997temperature}
T.~Iwashima, A.~Inoue, M.~Shigematsu, \emph{et~al.}, \enquote{Temperature compensation technique for fibre {B}ragg gratings using liquid crystalline polymer tubes,} {\protect\JournalTitle{Electronics letters}} \textbf{33}, 417--419 (1997).

\bibitem{huang2003temperature}
Y.~Huang, J.~Li, G.~Kai, \emph{et~al.}, \enquote{Temperature compensation package for fiber {B}ragg gratings,} {\protect\JournalTitle{Microwave and Optical Technology Letters}} \textbf{39}, 70--72 (2003).

\bibitem{Alxenses}
{Alxenses Company Ltd.}, \enquote{{Athermal Packaged FBG},} \url{http://www.alxenses.com/athermal_packaged_FBG.html}. Accessed on February 10, 2025.

\bibitem{Technicasa}
{Technica Optical Components, LLC.}, \enquote{{TWR50 Athermal FBG},} \url{https://technicasa.com/twr-50-athermal-fbg/}. Accessed on February 10, 2025.

\bibitem{Findlight}
{FindLight, LLC}, \enquote{{Athermal Packaged FBG},} \url{https://www.findlight.net/fiber-optics/fiber-optomechanics/fiber-bragg-grating/athermal-packaged-fbg/}. Accessed on February 10, 2025.

\bibitem{Alvarez}
C.~R. Alvarez, A.~Rahman, H.~{\"O}nel, \emph{et~al.}, \enquote{{Athermal package for OH suppression filters in astronomy: Part 1. Design},} in \emph{Advances in Optical and Mechanical Technologies for Telescopes and Instrumentation VI,}  vol. 13100 R.~Navarro and R.~Jedamzik, eds., International Society for Optics and Photonics (SPIE, 2024), p. 131002H.

\bibitem{Patent}
C.~{Rodriguez}, A.~{Rahman}, H.~{\"{O}nel}, \emph{et~al.}, \enquote{{Optisches Filtersystem zur Wellenlängenfilterung},} Patent application filed (20 June 2024).

\bibitem{2010SPIE.7735E..13Q}
A.~{Quirrenbach}, P.~J. {Amado}, H.~{Mandel}, \emph{et~al.}, \enquote{{CARMENES: Calar Alto high-resolution search for M dwarfs with exo-earths with a near-infrared Echelle spectrograph},} in \emph{Ground-based and Airborne Instrumentation for Astronomy III,}  vol. 7735 of \emph{Society of Photo-Optical Instrumentation Engineers (SPIE) Conference Series} I.~S. {McLean}, S.~K. {Ramsay}, and H.~{Takami}, eds. (2010), p. 773513.

\bibitem{2007Obs...127..171G}
R.~F. {Griffin}, \enquote{{Spectroscopic binary orbits from photoelectric radial velocities - Paper 194: HD 113997, HD 114931, HD 115588, and HD 116880, with a preliminary discussion of HD 113995},} {\protect\JournalTitle{The Observatory}} \textbf{127}, 171--188 (2007).

\bibitem{2024AJ....168..106R}
S.~A. {Rafi}, S.~K. {Nugroho}, M.~{Tamura}, \emph{et~al.}, \enquote{{Evidence of Water Vapor in the Atmosphere of a Metal-rich Hot Saturn with High-resolution Transmission Spectroscopy},} {\protect\JournalTitle{\aj}} \textbf{168}, 106--123 (2024).

\bibitem{2024A&A...691A.283B}
F.~{Biassoni}, F.~{Borsa}, F.~{Haardt}, and M.~{Rainer}, \enquote{{High-resolution transmission spectroscopy of the hot-Saturn HD 149026b},} {\protect\JournalTitle{\aap}} \textbf{691}, A283--A296 (2024).

\bibitem{2020A&A...638A..85R}
V.~{Roccatagliata}, E.~{Franciosini}, G.~G. {Sacco}, \emph{et~al.}, \enquote{{A 3D view of the Taurus star-forming region by Gaia and Herschel. Multiple populations related to the filamentary molecular cloud},} {\protect\JournalTitle{\aap}} \textbf{638}, A85--A95 (2020).

\bibitem{2021ApJ...919...55I}
T.~{Ichikawa}, M.~{Kido}, D.~{Takaishi}, \emph{et~al.}, \enquote{{Misaligned Circumstellar Disks and Orbital Motion of the Young Binary XZ Tau},} {\protect\JournalTitle{\apj}} \textbf{919}, 55--69 (2021).

\bibitem{2000A&A...354.1134R}
P.~{Rousselot}, C.~{Lidman}, J.~G. {Cuby}, \emph{et~al.}, \enquote{{Night-sky spectral atlas of OH emission lines in the near-infrared},} {\protect\JournalTitle{\aap}} \textbf{354}, 1134--1150 (2000).

\bibitem{Lachance}
R.~L. {Lachance}, A.~V. {Van}, M.~{Morin}, \emph{et~al.}, \enquote{{Adjustable athermal package for optical fiber devices},} U.S.patent 6,907,164B2 (14 June 2005).

\bibitem{proprieties_alu}
{Kloeckner Metals Germany GmbH}, \enquote{En aw-6082 3.2315,} \url{https://facts.kloeckner.de/werkstoffe/aluminium/3-2315/}. Accessed on February 10, 2025.

\bibitem{laser_coherent}
{Coherent, Inc}, \enquote{datasheet lasers innova-sabre-motofred,} \url{https://www.coherent.com/resources/datasheet/lasers/Innova-Sabre-MotoFreD-Data-Sheet.pdf}. Accessed on February 10, 2025.

\bibitem{hill1993bragg}
K.~O. Hill, B.~Malo, F.~Bilodeau, \emph{et~al.}, \enquote{Bragg gratings fabricated in monomode photosensitive optical fiber by {UV} exposure through a phase mask,} {\protect\JournalTitle{Applied Physics Letters}} \textbf{62}, 1035--1037 (1993).

\bibitem{montero2014influence}
A.~Montero, G.~Aldabaldetreku, G.~Durana, \emph{et~al.}, \enquote{Influence of humidity on fiber {B}ragg grating sensors,} {\protect\JournalTitle{Advances in Materials Science and Engineering}} \textbf{2014}, 405250--405258 (2014).

\bibitem{de1992effects}
B.~De~N{\`e}ve and M.~Shanahan, \enquote{Effects of humidity on an epoxy adhesive,} {\protect\JournalTitle{International Journal of Adhesion and Adhesives}} \textbf{12}, 191--196 (1992).

\bibitem{Wilson2002}
A.~J.~C. Wilson, \enquote{{The thermal expansion of aluminium from \SI{0}{^{\circ}C} to \SI{650}{^{\circ}C}},} {\protect\JournalTitle{Proceedings of the Physical Society}} \textbf{53}, 235--244 (2002).

\end{thebibliography}


%
%
%
%

\end{document}